\documentclass[twocolumn,aps]{revtex4}
\usepackage{amsmath}
\usepackage{graphicx}
\usepackage{dcolumn}
\usepackage{bm}

\begin{document}

\title{Quantum Monte Carlo study of a mag\-net\-ic-field-driv\-en 2D
su\-per\-con\-duc\-tor-in\-su\-la\-tor transition}
\author{Kwangmoo Kim and David Stroud}
\affiliation{Department of Physics, The Ohio State University, Columbus,
Ohio 43210, USA}
\date{\today}
\begin{abstract}


We numerically study the su\-per\-con\-duc\-tor-in\-su\-la\-tor phase transition
in a model disordered 2D superconductor as a function of applied magnetic field.
The calculation involves quantum Monte Carlo calculations of the $(2+1)$D $XY$
model in the presence of both disorder and magnetic field. The $XY$ coupling is
assumed to have the form $-J\cos(\theta_i-\theta_j-A_{ij})$, where $A_{ij}$ has a
mean of zero and a standard deviation $\Delta A_{ij}$. In a real system, such a
model would be approximately realized by a 2D array of small
Jo\-seph\-son-cou\-pled grains with slight spatial disorder and a uniform applied
magnetic field.  The different values $\Delta A_{ij}$ then corresponds to an applied
field such that the average number of flux quanta per plaquette has various integer
values $N$: larger $N$ corresponds to larger $\Delta A_{ij}$. For any value of
$\Delta A_{ij}$, there appears to be a critical coupling constant
$K_{c}(\Delta A_{ij}) = \sqrt{[J/(2U)]_c}$,
where $U$ is the charging energy, above which the system is a Mott insulator; there
is also a corresponding critical conductivity $\sigma^{*}(\Delta A_{ij})$ at
the transition.  For $\Delta A_{ij} = \infty$, the order parameter of the transition
is a renormalized coupling constant $g$.  Using a numerical technique appropriate
for disordered systems, we show that the transition at this value of $\Delta A_{ij}$
takes place from an insulating (I) phase to a Bose glass (BG) phase, and that the 
dynamical critical exponent characterizing this transition is $z \sim 1.3$. 
By contrast, $z = 1$ for this model at $\Delta A_{ij} = 0$.  We suggest that the 
superconductor to insulator transition is actually of this I to BG class at all 
nonzero $\Delta A_{ij}$'s, and we support this interpretation by both numerical 
evidence and an analytical argument based on the Harris criterion [A. B. Harris, 
J. Phys.\ C: Solid State Phys.\ {\bf 7}, 1671 (1974)]. 
$K_c$ is found to be a monotonically increasing function of $\Delta A_{ij}$.
For certain values of $K$, a disordered Josephson array may undergo a transition
from an ordered, Bose glass phase to an insulator with increasing $\Delta A_{ij}$.

\end{abstract}


\maketitle

\section{\label{sec:level1}INTRODUCTION}

The su\-per\-con\-duc\-tor-in\-su\-la\-tor (S-I) transition of thin
two-di\-men\-sion\-al (2D) superconducting films has been extensively studied both
theoretically \cite{cha1,cha2,fisher89,krauth,runge,sorensen,kampf,batrouni93,
makivic,wallin,hebert,schmid,lee1,lee2,nishiyama,capriotti,smakov} and
experimentally \cite{haviland,samband,markovic1,markovic2,yazdani} for many years.
The theoretical work can be broadly categorized into two groups: in one group,
disorder is induced using a random chemical potential, while in the other, disorder
is generated using a magnetic field. Most previous work belongs to the
former \cite{cha1,krauth,runge,sorensen,kampf,batrouni93,makivic,wallin,lee1,lee2,
smakov} whereas only a few belong to the latter \cite{cha2,nishiyama,markovic1,
markovic2}.

The present work is motivated primarily by several experiments in which an S-I
transition is observed in a 2D material as a function of applied transverse
magnetic field. Such experiments have been reported in thin films of
superconducting materials. They have also been carried out in some of the most
anisotropic cuprate high-$T_c$ superconductors; in such materials, individual
copper oxide layers may conceivably behave like thin superconducting films if they
are well enough decoupled from the other layers \cite{hebard,okuma95,goldman95,
kapri,okuma98,gantmakher00,markovic3,hao,gantmakher03,baturina04,steiner1,
baturina05,steiner2,aubin}. 
In both cases, the films seem to undergo a transition from S to I with increasing
magnetic field. Furthermore, the transition appears to be controlled mainly by the
film resistance $R$. Experiments suggest that, in contrast to some predictions, $R$
does not have a universal value at the S-I transition \cite{gantmakher00,markovic3}. 
In view of these experiments, it seems useful to construct a simple model which
contains disorder and also shows a field-driv\-en transition. In the present paper we
present such a model, and analyze its properties by a combination of numerical
methods and scaling assumptions.

Before describing our own approach, we briefly review some of the previous
theoretical work in this area. An early numerical calculation was carried out by
Cha {\it et al.}\ \cite{cha1} at zero magnetic field ($B = 0$). These workers
calculated both analytically and numerically the ze\-ro-tem\-per\-a\-ture ($T = 0$)
universal conductivity $\sigma^*$ of the 2D bos\-on-Hub\-bard model without
disorder at the S-I transition. They found, using numerical Monte Carlo
(MC) simulations of a (2+1)D $XY$ model, that
$\sigma^{*}=(0.285\pm 0.02)\sigma_{Q}$, where $\sigma_{Q}=(2e)^2/h$ is the quantum
conductance. This result is close to the value obtained from an analytic large-$N$
expansion. They further studied this model under an applied transverse magnetic
field using MC simulations, and found that $\sigma^{*}$ was increased \cite{cha2}.

Fisher {\it et al.}\ \cite{fisher89} studied the $T = 0$ phase diagrams and phase
transitions of bosons with short-range repulsive interactions moving in both
periodic and random potentials. For the periodic case, they found the system
exhibited two different phases, a superfluid and Mott insulator, and that the
dynamic exponent $z$ exactly equaled the spatial dimension $d$. They also derived
certain zero temperature constraints on the correlation length exponent $\nu$ and
the order parameter correlation exponent $\eta$, namely $\nu\geq 2/d$ and
$\eta\leq 2-d$. In the presence of disorder, they found that a ``Bose glass'' phase
existed, and that the transition to a superfluid phase took place from the Bose
glass phase, not directly from the Mott insulator.

Most previous studies in this area have been based on quantum Monte Carlo (QMC)
simulations \cite{krauth,runge,sorensen,batrouni93,makivic,wallin,hebert,schmid,
lee1,lee2,capriotti}.  Some work has involved advanced QMC techniques, such as a
QMC algorithm based on the exact duality transformation of the boson Hubbard
model \cite{hebert}, and a worm algorithm \cite{lee2}.  Other studies have used a
stochastic series expansion method \cite{hebert,smakov} and an exact diagonalization
method \cite{nishiyama}. Analytically, besides the large-$N$ expansion technique
used in Ref.\ \cite{cha1}, a coarse-grain\-ing approximation \cite{kampf} has been
adopted in some investigations.

The numerical studies of the S-I transition have used a wide range of model
Hamiltonians. Some workers have employed a 2D hard core boson
model \cite{runge,makivic,hebert,schmid,nishiyama}, while others used a 2D soft core
boson Hamiltonian \cite{fisher89,batrouni90,krauth,cha1}. This model has been used
to investigate the S-I transition at $T = 0$ \cite{sorensen,kampf,makivic,nishiyama,
smakov,samband}, as well as the su\-per\-con\-duc\-tor-Bo\-se glass (S-BG) phase
transition \cite{runge,batrouni93}, while some workers have investigated
both \cite{krauth,wallin,lee1,lee2}.  In addition, \v{S}makov and
S\o rensen \cite{smakov} studied the S-I transition at finite temperature $T$
using a similar model.

A number of workers have also investigated more complex phase transitions, of which
we mention just a few representative examples.
Capriotti {\it et al.}\ \cite{capriotti} studied a reentrant
su\-per\-con\-duc\-ting-to-nor\-mal (S-N) phase transition using, as a model, a
resistively shunted 2D Josephson junction array with normal Ohmic shunt resistors
as the source of dissipation. Chakravarty {\it et al.}\ \cite{chakravarty} also found
a dis\-si\-pa\-tion-in\-duced phase transition in such an array, but did not study
the possibility of reentrance. The reentrant S-N phase transition in
Ref.\ \cite{capriotti} was found to persist for moderate dissipation strength, but
the superconducting phase was always found to be stabilized above a critical
dissipation strength at sufficiently low $T$. H\'{e}bert {\it et al.}\ \cite{hebert}
studied phase transitions between superfluid, checkerboard, and striped solid
order, using two interactions---a near\-est-neigh\-bor ($V_1$) and
next-near\-est-neigh\-bor ($V_2$) repulsion---instead of a single parameter to
describe the random chemical potential. They found that the model exhibited a
superfluid to striped solid transition at half filling; away from half filling,
they found a first-order transition from superfluid to striped supersolid, as
well as a continuous transition from striped supersolid (superconducting) to
striped solid (insulating). Schmid {\it et al.}\ \cite{schmid} have studied a first
order transition between a checkerboard solid and a superfluid phase at finite
temperature. They also found that an unusual reentrant behavior in which ordering
occurs with increasing temperature. As an effort to develop a more realistic model,
several workers have included both short and long-range repulsive interactions
between bosons \cite{sorensen,wallin}, and some studies have included fluctuations
in the amplitude as well as the phase of the superconducting order
parameter \cite{kampf}.

The $T = 0$ S-I transition has been found to be characterized by universal
behavior. Ref.\ \cite{makivic} found, using QMC, that the dynamic exponent, the
correlation length exponent, and the universal conductivity were $z=0.5\pm 0.1$,
$\nu=2.2\pm 0.2$, and $\sigma_c=(1.2\pm 0.2)\sigma_{Q}$, respectively. In the
coarse-grain\-ing approximation \cite{kampf}, the universal conductivity was found
to be $\sigma^{*}=(\pi/8)\sigma_{Q}$, while the value
$\sigma^{*}=(0.45\pm 0.05)\sigma_{Q}$ was obtained at finite $T$ using the
stochastic series expansion with a geometric worm algorithm \cite{smakov}; in the
latter work it was also found that $\sigma/\sigma_Q$ scaled with $\omega/T$ at
small frequencies $\omega$ and low $T$.  
With only short-range Coulomb interactions, the universal conductivity at the phase
transition was found to be $\sigma^{*}=(0.14\pm 0.03)\sigma_{Q}$ \cite{sorensen,
wallin}.  With long-range Coulomb interactions, this value increased to
$\sigma^{*}=(0.55\pm 0.1)\sigma_{Q}$ \cite{sorensen} or
$\sigma^{*}=(0.55\pm 0.06)\sigma_{Q}$ \cite{wallin}.

This critical behavior differs significantly from the $T = 0$ S-BG transition. At
this transition, the dynamical exponent and the universal conductivity were found
to equal $z=1.95\pm 0.25$ and $\sigma^{*}=(0.17\pm 0.01)\sigma_{Q}$,
respectively \cite{runge}. Batrouni {\it et al.}\ \cite{batrouni93} found
$\sigma^{*}=(0.45\pm 0.07)\sigma_{Q}$ from QMC calculations and
$\sigma^{*}=(0.47\pm 0.08)\sigma_{Q}$ from analysis of cur\-rent-cur\-rent
correlation functions.

At intermediate strength of disorder, Lee {\it et al.}\ \cite{lee1} found that the
dynamical and the correlation length critical exponent were $1.35\pm 0.05$ and
$\nu=0.67\pm 0.03$, respectively. They also found that a Mott insulator to
superfluid transition occurred in the weak disorder regime while a Bose glass to
superfluid transition took place in the strong disorder regime.
More recently, Lee and Cha \cite{lee2} studied the quasiparticle energy gap near
the quantum phase transition. They found that this gap vanished discontinuously at
the transition for a weak disorder, implying a direct Mott insulator to
superfluid transition, whereas this discontinuous jump disappeared for a strong
disorder, supporting the intervention of Bose glass phase in this regime.

Several workers have studied the S-I transition by explicitly introducing a
magnetic field, using various models and experiments. For example,
Nishiyama \cite{nishiyama} found that the 2D hard core boson model exhibited a
field-tuned localization transition at a certain critical magnetic field and that
the critical DC conductivity was substantially larger than that at zero magnetic
field. In his work, the critical conductivity was found to be non-uni\-ver\-sal
but instead increased with increasing magnetic field.
Besides the experiments mentioned earlier,
Sambandamurthy {\it et al.}\ \cite{samband} found, from studies of thin amorphous
InO films near the S-I transition, that the resistivity followed a power law
dependence on the magnetic field in both the superconducting and the insulating
phases.

In most of the above calculations, the QMC approach is based on a mapping between a
$d$-di\-men\-sion\-al quantum mechanical system and a $(d+1)$-di\-men\-sion\-al
classical system with the imaginary time as an extra dimension \cite{sondhi}.  This
mapping works because calculating the thermodynamic variables of the quantum system
is equivalent to calculating the transition amplitudes of the classical system when
they evolve in the imaginary time. The imaginary time interval is fixed by the
temperature of the system. The net transition amplitude between two states of the
system can then be obtained by a summation over the amplitudes of all possible
paths between them according to the prescription of Feynman \cite{feynman}.  These
paths are the states of the system at each intermediate time step. Therefore, the
path-in\-te\-gral description of the quantum system can be interpreted using the
statistical mechanics of the $(d + 1)$-di\-men\-sion\-al classical system held at
a fictitious temperature which measures ze\-ro-point fluctuations in the quantum
system.

In order for a boson system to have a su\-per\-con\-duc\-tor-in\-su\-la\-tor
transition, the bosons must have an on-site repulsive interaction, i.e., a
``charging energy.'' Otherwise the boson system would usually undergo
Bo\-se-Ein\-stein condensation at zero temperature.
The charging energy induces ze\-ro-point fluctuations of the phases and disorders
the system. On the other hand the Josephson or $XY$ coupling favors coherent
ordering of the phases, which causes the onset of superconductivity. Therefore, the
competition between the charging energy and $XY$ coupling is responsible for the
su\-per\-con\-duc\-tor-in\-su\-la\-tor transition, which typically occurs at a
critical value of the ratio of the strengths of these two energies. In addition, if
a disorder is added to the system, the system may also undergo a transition to a
phase other than a Mott insulator, depending on the strength of the disorder. This
additional phase is known to be a Bose glass phase.

In this work, we study the ze\-ro-tem\-per\-a\-ture quantum phase transitions of 2D
model superconducting films in an applied magnetic field. Our model includes both
charging energy and Josephson coupling, and thus allows for an S-I transition. In
our approach, the applied magnetic field is described by a root-mean-square (rms) 
fluctuation $\Delta A_{ij}$ which describes the randomness in the flux per plaquette. 
This randomness leads to the occurrence of a Bose glass phase at large $\Delta A_{ij}$.
As explained further below, our model corresponds well to a 2D Josephson junction
array with weak disorder in the plaquette areas, studied at an applied uniform
magnetic field corresponding to integer number $N_v$, on average, of flux quanta
per plaquette. The quantity $\Delta A_{ij}$ is proportional to the rms disorder in 
the flux per plaquette, and is proportional to $N_v$. Thus, our model gives rise to 
an S-I transition in the array with increasing $N_v$ (or increasing magnetic field). 

The remainder of this paper is organized as follows. Sec.\ \ref{sec:level2}
presents the formalism. In this section, we give the model boson Hubbard model,
and describe its conversion to a $(2+1)$D $XY$ model, which we treat using
path-in\-te\-gral Monte Carlo calculations. We also describe the fi\-nite-size
scaling methods for obtaining the critical coupling constants and universal
conductivities at the transition. Finally, this section describes the nature of
the renormalized coupling constant used to study the behavior of the system in
the fully random case. Sec.\ \ref{sec:level3} presents our numerical results,
using these approaches. We discuss our results and present our conclusions in
Sec.\ \ref{sec:level4}.

\section{\label{sec:level2}FORMALISM}

\subsection{Model Hamiltonian}

Our goal is to examine the su\-per\-con\-duc\-tor-in\-su\-la\-tor transition in a
disordered 2D system in a magnetic field at very low temperature $T$. Thus, a
useful model for this transition would include three features: (i) a competition
between a Coulomb energy and an energy describing the hopping of Cooper pairs;
(ii) disorder; and (iii) a magnetic field. In particular, we hope that this model
will exhibit, for suitable parameters, a transition from superconductor to
insulator as the magnetic field is increased. While there are a wide range of
models which could incorporate these features, we choose to consider a model
Hamiltonian appropriate to a 2D Josephson junction array:
\begin{equation}
{\cal H}^\prime = U\sum_jn_j^2 - J\sum_{\langle ij\rangle}
\cos(\theta_i-\theta_j-A_{ij}).
\label{eq:calh}
\end{equation}
Here $n_j$ is the operator representing the number of Cooper pairs on a site $j$,
$J$ is the Josephson energy coupling sites $i$ and $j$, $\theta_i$ is the phase 
of the order parameter on the $i$th site, 
$A_{ij}=(2\pi/\Phi_{0})\int_{i}^{j}{\bf A}\cdot d{\bf l}$ is a magnetic phase
factor, $\Phi_{0}=hc/2e$ is the flux quantum, and ${\bf A}$ is the vector
potential. In this picture, each site can be thought of as a superconducting grain.

For calculational convenience, we choose to take the sites $j$ to lie on a
{\em regular} 2D lattice (a square lattice in our calculations), with Josephson
coupling only between nearest neighbors. Thus, the disorder in our model is
incorporated via the magnetic phase factors $A_{ij}$, as explained further below.
Our Hamiltonian is identical to that of Cha {\it et al.}\ \cite{cha1} except that
we consider the special case that the chemical potential $\mu_i$ for Cooper pairs
on the $i$th grain is an integer, and we choose the $A_{ij}$'s to be random.

The first term in Eq.\ (\ref{eq:calh}) is the charging energy. We consider only
a diagonal charging energy and also assume all grains to be of the same size,
so that $U$ is independent of $j$. Since the charging energy $E_{Cj}$ of a grain
carrying charge $Q_j$ with capacitance $C$ is $E_{Cj} = Q_j^2/(2C)$,
\begin{equation}
U = \frac{(2e)^2}{2C} = \frac{2e^2}{C}. 
\label{eq:u}
\end{equation}
We also know that $Q_j = CV_j$, where $V_j$ is the voltage of grain $j$ relative
to ground; so
\begin{equation}
E_{Cj} =\frac{1}{2}CV_j^2 = \frac{C\hbar^2}{2(2e)^2}\dot{\theta}_j^2,
\end{equation}
where we have used the Josephson relation, $V_j = (\hbar/2e)\dot{\theta}_j$.
Finally, we can express $C$ in terms of $U$ using Eq.\ (\ref{eq:u}), with the result
\begin{equation}
E_{Cj} = \frac{\hbar^2}{4U}\dot{\theta}_j^2. 
\label{eq:ec}
\end{equation}
Combining all these relations, we obtain
\begin{equation}
{\cal H}^\prime = \frac{\hbar^2}{4U}\sum_j\dot{\theta}_j^2 -
J\sum_{\langle ij \rangle}\cos(\theta_i - \theta_j - A_{ij}).
\label{eq:hamil}
\end{equation}

Since we have taken the grains to lie on a lattice, we need to choose the
$A_{ij}$'s in a way which incorporates randomness. Thus, we make the simplifying
assumption that the phase factor $A_{ij}$ of each bond in the plane is an
independent Gaussian random variable with a mean of zero and a standard deviation
$\Delta A_{ij}$:
\begin{equation}
P(A_{ij})=\frac{1}{\sqrt{2\pi}(\Delta A_{ij})}\exp\left[-\frac{A_{ij}^2}
{2(\Delta A_{ij})^2}\right].
\label{eq:pa}
\end{equation}
Since the sum of the phase factors around the four bonds of a plaquette is
$2\pi/\Phi_0$ times the flux through that plaquette, this choice will cause the
flux through the plaquette also to be a random variable. However, the fluxes
through near\-est-neigh\-bor plaquettes will be correlated.

\begin{figure}[ht!]
\begin{center}
\includegraphics[width=0.35\textwidth]{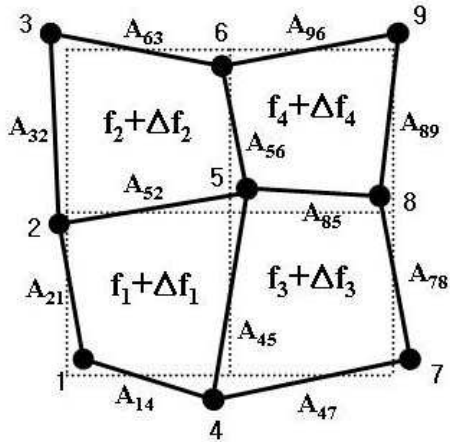}\\%
\end{center}
\caption{\label{fig:rmsflux}Sketch of a $2 \times 2$ group of plaquettes in a square
lattice of Jo\-seph\-son-cou\-pled grains, in which each grain is randomly
displaced by a small amount from its nominal lattice site. In a uniform transverse
magnetic field, if the average flux $\Phi$ per plaquette is $\Phi = f\Phi_0$, then
the root-mean-square deviation of the flux from its mean value is also proportional
to $f$.}
\end{figure}

Although this model may seem artificial, it should closely resemble a real,
physically achievable system. Specifically, consider a spatially {\it random}
distribution of grains in 2D in a {\it uniform} magnetic field. Suppose that the
grain positions deviate slightly (but randomly) from the sites of a square lattice.
Then the areas of the square plaquettes have a random distribution, and
consequently, the flux through each plaquette also varies randomly about its mean.
If the average flux $\Phi$ per plaquette is $\Phi = f\Phi_0$, then the
root-mean-square deviation of the flux, $\Delta\Phi \propto f$ as in
Fig.\ \ref{fig:rmsflux}.

Now consider the special case of integer $f$. In the absence of disorder, the 2D
array at integer $f$ should behave exactly like the array at $f = 0$, because the
Hamiltonian would then be perfectly periodic in $f$ \cite{shih1,shih2,shih3}.
With nonzero disorder, only the {\em rms deviation} from integer $f$, i.e.,
$\Delta f = \Delta\Phi/\Phi_0$, is physically relevant. This deviation increases
linearly with $f$.

In short, our model Hamiltonian is approximately realized by a 2D Josephson
junction array on a square lattice, in which the grains deviate randomly in
position from their lattice sites, placed in a transverse magnetic field with an
average flux per plaquette $f\Phi_0$, with integer $f$. A larger $\Delta A_{ij}$
corresponds to a larger $f$. The models are not equivalent, even if the random
position deviations are specified by Gaussian variables, because we assume the
$\Delta A_{ij}$'s for different bonds are uncorrelated, whereas they would be
correlated in the positionally disordered case. However, this difference should
have little effect in practice and we have confirmed this in the limit of large
$f$ (see below). For non-in\-te\-ger $f$, the disordered 2D Josephson array has
well-known oscillatory properties as a function of $f$ which are not described
by our model as formulated above. This disordered Josephson array is not an
entirely realistic model of a superconducting {\it film} which undergoes a
field-driv\-en su\-per\-con\-duc\-tor-in\-su\-la\-tor transition because our model
involves an underlying lattice of grains. Nonetheless, we may hope that some of
the properties of our model resemble those seen in experimentally studied materials.

A related method of including random flux has previously been used by Huse and
Seung \cite{huse} as a model for a three-di\-men\-sion\-al (3D) ``gauge glass.''
These workers considered only $\Delta A_{ij} = \infty$, and studied a 3D classical
model (U = 0) rather than the 2D quantum case considered here.

\subsection{Path Integral Formulation}

We can now use the model Hamiltonian (\ref{eq:hamil}) to obtain the action
in the form of a standard integral over imaginary time. The action $S$ may be
written
\begin{equation}
\frac{S}{\hbar} = \frac{1}{\hbar}\int {\cal L}d\tau,
\end{equation}
where ${\cal L}$ is the Lagrangian, given by
\begin{equation}
{\cal L } =
\frac{\hbar^2}{4U}\sum_j\left(\frac{\partial\theta_j}{\partial\tau}\right)^2
- J\sum_{\langle ij \rangle}\cos[\theta_i(\tau) - \theta_j(\tau) - A_{ij}(\tau)].
\end{equation}

The partition function is now given by a path integral of $\exp(-S/\hbar)$
over all possible paths described by the variables $\theta_i(\tau)$ in
imaginary time $\tau$, integrated from $\tau = 0$ to $\tau = \beta\hbar$,
where $\beta = 1/(k_BT)$. This path integral can be reduced to the partition
function of an anisotropic classical $XY$ model in three dimensions. Here, by
``anisotropic'' we mean that the coupling constants $K$ and $K_\tau$ in the $xy$
plane and $\tau$ direction are different. To make the mapping, we first write
\begin{equation}
\left(\frac{\partial\theta_i}{\partial\tau}\right)^2 \sim
\left(\frac{\Delta\theta_i}{\Delta\tau}\right)^2 \sim
\frac{2 - 2\cos\Delta\theta_i}{(\Delta \tau)^2},
\end{equation}
where $\Delta\tau$ is the width of the time slice, $\Delta\theta_i =
\theta_i(\tau + \Delta\tau)-\theta_i(\tau)$, and we have used the expansion
of $\cos\Delta\theta$ to second order in the small quantity $\Delta\theta$.
This expansion is accurate when $\Delta \tau$ is sufficiently small.

Neglecting the constant term in this expansion, we finally obtain
\begin{eqnarray}
\frac{S}{\hbar} & = & - K_\tau\sum \cos[\theta_i(\tau) - \theta_i(\tau +
\Delta\tau)] \nonumber \\ & & - K\sum\cos[\theta_i(\tau) - \theta_j(\tau) - A_{ij}(\tau)].
\end{eqnarray}
Here the sums run over all bonds in the $\tau$ direction and in the $xy$
plane, respectively.

In order to obtain the values of the coupling constants $K$ and $K_\tau$,
we assume that we have broken up the time integral into $M$ time slices,
each of width $\beta\hbar/M$. Then the coupling constant in the $xy$ direction
is just
\begin{equation}
K = \frac{\beta J}{M}.
\end{equation}
The coupling constant in the $\tau$ direction is given by
\begin{equation}
K_\tau =
\frac{\hbar^2}{4U}\frac{1}{\hbar}(\Delta\tau)\frac{2}{(\Delta\tau)^2}
= \frac{1}{2U}\frac{\hbar}{\Delta\tau} = \frac{M}{2\beta U},
\end{equation}
we have used $\Delta\tau = \beta\hbar/M$ and included the extra factor
of $2$ in the numerator because $\cos\Delta\theta \sim 1 - (\Delta\theta)^2/2$.

Given $K$ and $K_\tau$, the partition function is obtained from this
anisotropic 3D $XY$ classical Hamiltonian with coupling constants
$K$ and $K_\tau$. Any desired equilibrium quantity can, in principle,
be computed by averaging over all configurations using standard classical
Monte Carlo techniques. Within any given realization of the disorder,
the $A_{ij}$'s are chosen at random from the Gaussian distribution
within the $xy$ plane, as described above, but the $A_{ij}$'s for a given
bond in the $xy$ plane are independent of $\tau$, i.e., they are the same
for all time slices. In principle, for any given $\beta$, this should be
done taking the limit as $M \rightarrow \infty$. In practice, of course,
the size of the sample is limited by considerations of computer time.

\subsection{Evaluation of Specific Properties Using Path Integral Formulation}

The time-slice formulation of the partition function allows various
properties to be evaluated using standard classical Monte Carlo
techniques. We now review how this may be done for the helicity modulus
(or superfluid density) and the electrical conductivity. Similar
formulations have been given in Refs.\ \cite{fisher73,ohta,shih2,shih3,cha1,cha2}
for different but related models.

\subsubsection{Helicity Modulus}

For a frustrated classical $XY$ system in $d$ dimensions, the helicity
modulus tensor $\gamma_{\alpha\beta}$ is a $d \times d$ matrix which is
a measure of the phase stiffness. It is defined as the second derivative
of the free energy with respect to an infinitesimal phase twist, and may
be written
\begin{equation}
\gamma_{\alpha \beta}=\frac{1}{N}\left.
\frac{\partial^{2}F}{\partial A_{\alpha}^\prime \partial A_{\beta}^\prime}
\right |_{{\bf A}^\prime=0},
\label{eq:twist}
\end{equation}
where $N$ is the number of sites in the system and ${\bf A}^\prime$ is
a fictitious vector potential added to the Hamiltonian (in addition to
the vector potential ${\bf A}$ already included in the
Hamiltonian \cite{fisher73}). In explicit form, this derivative takes
the following form for the diagonal elements (see, e.g., Ref.\ \cite{kim}):
\begin{eqnarray}
& & \gamma_{\alpha\alpha} = \frac{1}{N}\left\langle\sum_{\langle ij
\rangle}J_{ij}\cos(\theta_i-\theta_j -
A_{ij})(\hat{e}_{ij}\cdot\hat{e}_\alpha)^2\right\rangle \nonumber \\
& & - \frac{1}{Nk_BT}\left\langle\left[\sum_{\langle ij
\rangle}J_{ij}\sin(\theta_i-\theta_j -
A_{ij})(\hat{e}_{ij}\cdot\hat{e}_\alpha)\right]^2\right\rangle
\nonumber \\
& & +\frac{1}{Nk_BT}\left\langle\sum_{\langle ij
\rangle}J_{ij}\sin(\theta_i-\theta_j -
A_{ij})(\hat{e}_{ij}\cdot\hat{e}_\alpha)\right\rangle^2.
\end{eqnarray}
Here $\hat{e}_{ij}$ is a unit vector from the $i$th to the $j$th site,
and $\hat{e}_\alpha$ is a unit vector in the $\alpha$ direction. The
triangular brackets denote an average in the canonical ensemble.

If this expression is applied to the time-slice representation of the
quan\-tum-me\-chan\-i\-cal Hamiltonian, the coupling constants $J_{ij}$ will
be different in the $xy$ plane and in the $\tau$ direction. For the
time-slice calculation, we have to be careful in order to obtain a
result which is well-be\-haved in the limit $M \rightarrow \infty$,
where $M$ is the number of time slices. The correct expression in
this case is
\begin{eqnarray}
& & \gamma_{xx} = \frac{1}{N_xN_y}\left\langle\frac{J}{M}\sum_{\langle
ij\rangle \| \hat{x}}\cos(\theta_i - \theta_j - A_{ij})\right\rangle
\nonumber \\
& & - \frac{1}{N_xN_yk_BT}\left\langle \left[\sum_{\langle ij \rangle \|
\hat{x}}\frac{J}{M}\sin(\theta_i - \theta_j - A_{ij})\right]^2\right\rangle
\nonumber \\
& & + \frac{1}{N_xN_yk_BT}\left\langle \sum_{\langle ij \rangle \|
\hat{x}}\frac{J}{M}\sin(\theta_i - \theta_j - A_{ij})\right\rangle^2.
\label{eq:gamx}
\end{eqnarray}
Here we are assuming that there are $N_xN_y$ superconducting grains in
our 2D lattice and $M$ time slices. The sums run over all distinct bonds
in the $\hat{x}$ direction; there are $N_xN_yM$ of these bonds ($N_xN_y$
per time slice). A similar expression holds for $\gamma_{yy}$. The
in-plane coupling constant is taken to be $J/M$ because there are $M$
time slices.

From the above form, we can see why the expression behaves correctly in
the limit $M \rightarrow \infty$. Each of the two sums contains $N_xN_yM$
terms in it, but the ensemble average consists of $M$ identical terms, one
for each layer. Therefore, the first sum in Eq.\ (\ref{eq:gamx}), for example,
should take the form
\begin{eqnarray}
& & \frac{J}{M}\left\langle\sum_{\langle ij
\rangle}\cos(\theta_i-\theta_j-A_{ij})\right\rangle \nonumber \\ & & 
\longrightarrow J\left\langle\sum^\prime_{\langle ij
\rangle}\cos(\theta_i-\theta_j-A_{ij})\right\rangle,
\end{eqnarray}
where the sum on the right hand side runs only over the phases in a
{\em single layer}. The right-hand side is evidently independent of $M$
in the limit $M \rightarrow \infty$. A similar argument can be used to
show that the second part of the expression (\ref{eq:gamx}) for
$\gamma_{xx}$ also approaches a well-be\-haved limit as $M \rightarrow
\infty$. Our numerical results confirm this behavior.

As an illustration, we write down an expression for $\gamma_{xx}$ in
the limit $T \rightarrow 0$ in the unfrustrated case ($\Delta A_{ij} = 0$).
First, we multiply expression (\ref{eq:gamx}) by $\beta/M$ to obtain
\begin{eqnarray}
& & \frac{\beta\gamma_{xx}}{M} = \frac{1}{N_xN_yM}\left\langle
K\sum_{\langle ij \rangle \| \hat{x}}\cos(\theta_i-\theta_j-A_{ij})
\right\rangle \nonumber \\
& & -\frac{1}{N_xN_yM}\left\langle\left[\sum_{\langle ij \rangle \|
\hat{x}}K\sin(\theta_i-\theta_j-A_{ij})\right]^2\right\rangle
\nonumber \\
& & +\frac{1}{N_xN_yM}\left\langle\sum_{\langle ij \rangle \|
\hat{x}}K\sin(\theta_i-\theta_j-A_{ij})\right\rangle^2,
\end{eqnarray}
where $K = \beta J/M$. The corresponding coupling constant in the
$\tau$ direction is $K_\tau = M/(2\beta U)$.

Since we are interested in the limit $\beta \gg 1$, we choose $M$
so that $K = K_\tau$. This condition is equivalent to
\begin{equation}
\frac{\beta}{M} = \frac{1}{\sqrt{2JU}}.
\end{equation}
Hence, we get
\begin{eqnarray}
& & \frac{\gamma_{xx}}{\sqrt{2JU}} = \frac{1}{N_xN_yM}\left\langle
K\sum_{\langle ij \rangle \|
\hat{x}}\cos(\theta_i-\theta_j-A_{ij})\right\rangle \nonumber \\
& & - \frac{1}{N_xN_yM}\left\langle\left[\sum_{\langle ij \rangle \|
\hat{x}}K\sin(\theta_i-\theta_j-A_{ij})\right]^2\right\rangle \nonumber \\
& & + \frac{1}{N_xN_yM}\left\langle\sum_{\langle ij \rangle \|
\hat{x}}K\sin(\theta_i-\theta_j-A_{ij})\right\rangle^2,
\end{eqnarray}
where $K = \beta J/M = \sqrt{J/(2U)}$. Since $K = K_\tau$, the
right-hand side of this equation represents a dimensionless helicity
modulus $\tilde{\gamma}$ for a classical un\texttt{frustrat}ed isotropic 3D $XY$
model on a simple cubic lattice, which is a function of a single dimensionless
coupling constant $K$\@.

Now it is known from previous Monte Carlo studies \cite{li,gottlob,fernandez,
schultka,ryu} that the unfrustrated 3D $XY$ model on a simple cubic lattice has an 
ordered phase if $K > K_c \sim 1/2.21 \sim 0.452$. Therefore, $\tilde{\gamma}(K)$ 
vanishes if $K < K_c$ and is positive for $K > K_c$. Translating this result to 
the 2D quantum $XY$ model on a square lattice, we see that there is a
su\-per\-con\-duc\-tor-in\-su\-la\-tor transition at $J/(2U) = (0.452)^2 = 0.204$.

For reference we give the connection between our formulation of the helicity
modulus and the calculation of Cha {\it et al.}\ \cite{cha1} Rather than the
helicity modulus, these workers calculate the quantity $\rho(0)$,
which is related to the superfluid density $\rho_s$ by
\begin{equation}
\rho(0) = \frac{\rho_s}{k_BT}
\end{equation}
and to the components of the helicity modulus tensor by
$\rho(0) = K\gamma \equiv K(\gamma_{xx}+\gamma_{yy})/2$, where
$\gamma_{xx} = \gamma_{yy}$ for the present model, which is isotropic in the
$xy$ plane. In our notation, $\rho(0)$ is given by
\begin{eqnarray}
\rho(0) & = & \frac{JK}{2N_xN_yM}\left[\left\langle \sum_{\langle ij \rangle \|
\hat{x}}\cos(\theta_i-\theta_j-A_{ij}) \right\rangle\right. \nonumber \\ 
& & \left. +\left\langle \sum_{\langle ij
\rangle \| \hat{y}} \cos(\theta_i-\theta_j-A_{ij}) \right\rangle\right] \nonumber \\
& & -\frac{JK^2}{2N_xN_yM}\left[\left\langle\left[\sum_{\langle ij \rangle \|
\hat{x}} \sin(\theta_i-\theta_j-A_{ij})\right]^{2}\right\rangle\right. \nonumber \\
& & \left. + \left\langle\left[\sum_{\langle ij \rangle \| \hat{y}}
\sin(\theta_i-\theta_j-A_{ij})\right]^{2}\right\rangle\right] \nonumber \\
& & +\frac{JK^2}{2N_xN_yM}\left[\left\langle\sum_{\langle ij \rangle \| \hat{x}}
\sin(\theta_i-\theta_j-A_{ij})\right\rangle^{2}\right. \nonumber \\
& & \left. + \left\langle\sum_{\langle ij \rangle \| \hat{y}}
\sin(\theta_i-\theta_j-A_{ij})\right\rangle^{2}\right].
\label{eq:rho0}
\end{eqnarray}

\subsubsection{Specific Heat}

For the specific heat $C_V$, we used the fluc\-tu\-a\-tion-dis\-si\-pa\-tion
theorem given by
\begin{equation}
C_{V}=\frac{\langle \mathcal{H}^{\prime 2} \rangle - \langle \mathcal{H}^\prime
\rangle^2}{Nk_{B}T^2},
\end{equation}
where $N$ is the total number of sites in the lattice, $\mathcal{H}^\prime$
is the Hamiltonian in Eq.\ (\ref{eq:hamil}), and $\langle \cdots \rangle$
denotes an ensemble average.

\subsubsection{Conductivity}

The conductivity in the low-fre\-quen\-cy limit can also be obtained from the
time-slice Monte Carlo approach as \cite{cha1}
\begin{equation}
\sigma(0) = 2\pi\sigma_Q\lim\limits_{k\to 0}\frac{\rho(k)}{k},
\label{eq:conduct}
\end{equation}
where $\rho(k)$ is proportional to the superfluid density at frequency $k$ and 
is given by
\begin{eqnarray}
\rho(k) & = & \frac{JK}{2N_xN_yM}\left[\left\langle \sum_{\langle ij \rangle \|
\hat{x}}\cos(\theta_i-\theta_j-A_{ij}) \right\rangle\right. \nonumber \\ 
& & \left. +\left\langle \sum_{\langle ij
\rangle \| \hat{y}}\cos(\theta_i-\theta_j-A_{ij}) \right\rangle\right] \nonumber \\
& & - \frac{JK^2}{2N_xN_yM}\left\langle\sum_{\langle ij \rangle \|\hat{x},{\bf x}}
\sin(\theta_i-\theta_j-A_{ij})e^{-i{\bf k}\cdot{\bf x}}\right. \nonumber \\ 
& & \left. \times\sum_{\langle ij \rangle \|
\hat{x},{\bf x}} \sin(\theta_i-\theta_j-A_{ij})e^{i{\bf k}\cdot{\bf x}}
\right\rangle \nonumber \\
& & - \frac{JK^2}{2N_xN_yM}\left\langle\sum_{\langle ij \rangle \|\hat{y},{\bf x}}
\sin(\theta_i-\theta_j-A_{ij})e^{-i{\bf k}\cdot{\bf x}}\right. \nonumber \\ 
& & \left. \times\sum_{\langle ij \rangle \|
\hat{y},{\bf x}} \sin(\theta_i-\theta_j-A_{ij})e^{i{\bf k}\cdot{\bf x}}
\right\rangle \nonumber \\
& & + \frac{JK^2}{2N_xN_yM}\left\langle\sum_{\langle ij \rangle \| \hat{x},{\bf x}}
\sin(\theta_i-\theta_j-A_{ij})e^{-i{\bf k}\cdot{\bf x}}\right\rangle \nonumber \\
& & \times\left\langle\sum_{\langle ij \rangle \| \hat{x},{\bf x}}
\sin(\theta_i-\theta_j-A_{ij})e^{i{\bf k}\cdot{\bf x}}\right\rangle \nonumber \\
& & + \frac{JK^2}{2N_xN_yM}\left\langle\sum_{\langle ij \rangle \| \hat{y},{\bf x}}
\sin(\theta_i-\theta_j-A_{ij})e^{-i{\bf k}\cdot{\bf x}}\right\rangle \nonumber \\
& & \times\left\langle\sum_{\langle ij \rangle \| \hat{y},{\bf x}}
\sin(\theta_i-\theta_j-A_{ij})e^{i{\bf k}\cdot{\bf x}}\right\rangle.
\label{eq:rhok}
\end{eqnarray}
In the limit of very small $k$, we expect that $\rho(k)$ will remain finite in
the superconducting phase and vanish in the insulating phase. Thus, $\sigma(0)$
will become infinite in the superconducting phase but vanish in the insulating
phase. Precisely at the critical value $K_c$, $\sigma(0)$ will become finite with
a universal value, as already obtained by other workers for related models.

\section{\label{sec:level3}Quantum Monte Carlo Results}

\subsection{Numerical Procedure}

In our quantum Monte Carlo calculations, we use the standard Metropolis algorithm
with periodic boundary conditions in both the spatial directions and the imaginary
time direction. We usually start with a random configuration of phases at $K=0.4$,
then increase $K$ up to $K=0.7$ in steps of $0.005$. This procedure corresponds to
lowering the temperature $T$ since $K\propto 1/T$. At each $K$, we take 40000 MC
steps per site through the entire lattice to equilibrate the system,
after which we take an additional 50000 MC steps to calculate the thermodynamic
variables of interest. For a lattice size of $6^3$, we use ten times as many MC
steps as these for both equilibration and averaging, and for a lattice size of
$8^3$, we use twice as many steps.

For the phases $\theta_i$ of the order parameter on each site, we use the
$360$-state clock model instead of a continuous angle between $0$ and $2\pi$ since
it allows us to cover the entire range of angles with fewer trials.
Therefore, the allowable angles are $0^{\circ}$, $1^{\circ}$, $2^{\circ}$,...,
$359^{\circ}$. It has been shown numerically that these discrete phase angles give
results indistinguishable from the continuous ones provided that
$n>20$ \cite{thijssen}. However, we select $A_{ij}$ from a continuous distribution 
in all our calculations.

For the partially random and completely random $A_{ij}$, we averaged over 100
different realizations of $\Delta A_{ij}$ to calculate the helicity modulus $\gamma$ and
the specific heat $C_V$. These calculations were so time-con\-sum\-ing that we
could go just up to $20\times 20\times 20$ lattice size.
For this reason, we chose to 
carry out simulations only over four different $\Delta A_{ij}$ for the partially
random case: $\Delta A_{ij}=1/2$, $1/\sqrt{2}$, $(1+1/\sqrt{2})/2\approx 0.854$,
and $1$.  
Each realization is specified by a different random number seed.  

\subsection{Fi\-nite-Size Scaling for $\gamma(0)$}

In general, if there is a continuous phase transition as a function of some
parameter, such as the coupling constant $K$, the critical behavior near the
transition can be analyzed by carrying out a fi\-nite-size scaling analysis of
various calculated quantities. For example, the zero-fre\-quen\-cy
helicity modulus $\gamma(0)$ is expected to satisfy \cite{wallin}
\begin{equation}
\gamma(0)= \frac{1}{L^{d+z-2}}\tilde{\gamma}\left (L^{1/\nu}\delta,\frac{L_{\tau}}
{L^z}\right ), \label{eq:gamma0}
\end{equation}
where $d$ is the spatial dimensionality, $z$ is the dynamic exponent,
$\tilde{\gamma}$ is a scaling function, $\nu$ is the critical exponent for the
correlation length $\xi$, $\delta = (K-K_c)/K_c$, $K_c$ is the critical value
of the coupling constant, and $L_{\tau}$ is the thickness in the imaginary
time direction. For our present system, $d = 2$, so the right-hand side is 
$L^{-z}\tilde{\gamma}(L^{1/\nu}\delta, L_\tau/L^z)$. If we define $\tilde{\gamma}
(L^{1/\nu}\delta, L_\tau/L^z)=(L^z/L_\tau)\tilde{G}(L^{1/\nu}\delta, L_\tau/L^z)$, 
then this scaling relation becomes
\begin{equation}
L_\tau\gamma(0) = \tilde{G}\left(L^{1/\nu}\delta, \frac{L_\tau}{L^z}\right).
\end{equation}
If our computational box has $N_x \times N_y \times M$ sites, with $N_x = N_y$,
this scaling relation may be equivalently written as
\begin{equation}
KM\gamma(0) = \tilde{G}\left(N_x^{1/\nu}\delta, \frac{M}{N_x^z}\right).
\label{eq:gamma0a}
\end{equation}

As has been noted by other workers \cite{cha1,sorensen,cha2,wallin,lee1,hlee, 
fabien,hitchcock}, $K_c$ can now be found, 
if $\gamma$ is a suitable order parameter, by plotting $KM\gamma(0)$ as a function
of $\delta$ for various cell sizes, all with aspect ratios satisfying $M = cN_x^z$,
and finding the point where these all cross, which corresponds to $\delta = 0$.
Unfortunately, this method requires knowing the value of $z$ in advance. For many
such quantum phase transitions, $z$ may not be known. Thus, one should carry out
this calculation for {\it all} plausible values of $z$ and find out which value
leads to a satisfactory crossing. This procedure is prohibitively demanding
numerically. We have, therefore, initially attempted to carry out 
scaling using $z = 1$, the value which is known to be correct at $\Delta A_{ij} = 0$.  
If this value were correct also at $\Delta A_{ij} \neq 0$, it would suggest that 
the su\-per\-con\-duc\-ting-in\-su\-la\-ting transition at finite $\Delta A_{ij}$ 
is in the same universality class as the ze\-ro-field transition. 
In practice, we find that $z = 1$ never gives perfect scaling at nonzero 
$\Delta A_{ij}$ and the scaling fit becomes progressively worse as $\Delta A_{ij}$ 
increases. For large $\Delta A_{ij}$ in particular, the fit clearly fails, and we 
find for these values that $\gamma(0)$ never converges to a nonzero value. 
At such large $\Delta A_{ij}$, we suggest that this regime corresponds to a Bose 
glass (as further discussed below), and carry out a different kind of scaling 
calculation to obtain the actual value of $z$ of this phase at 
$\Delta A_{ij} = \infty$.  We also give arguments suggesting that, in fact, this 
Bose glass phase is actually the ordered phase for {\em all} nonzero values of 
$\Delta A_{ij}$.


Operationally, we implement the hypothesis that $z = 1$ by taking $N_x=N_y=M$.
With this choice, Eq.\ (\ref{eq:gamma0a}) becomes
\begin{equation}
KM\gamma(0)=\tilde{G}(M^{1/\nu}\delta, 1)
\label{eq:gamma1}
\end{equation}
when $d=2$.

\subsection{Zero Magnetic Field}

\begin{figure}[ht!]
\begin{center}
\includegraphics[width=0.4\textwidth]{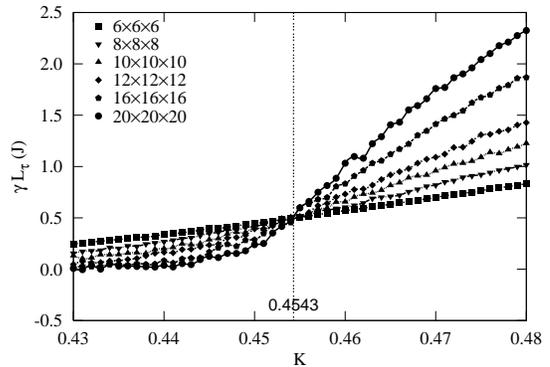}\\%
\end{center}
\caption{\label{fig:scaling_gamma}Plot of $\gamma L_\tau$
as a function of $K$ for various $N_x \times N_y \times M$ lattices for
$A_{ij} = 0$. In this and all subsequent figures, unless otherwise specified, we
use $N_x=N_y=M$. The phase transition occurs where the curves of different $M$
cross. The crossing point yields $K_c=0.4543\pm 0.0011$. The use of $N_x=N_y=M$
is equivalent to assuming that the dynamic exponent $z = 1$, as discussed in the
text.}
\end{figure}

As a check of our method, we have calculated $K_c$ for the case of zero magnetic
field ($A_{ij} = 0$), using the above numerical approach. When there is no magnetic
field, we get $K_c=0.4543\pm 0.0011$ using a fi\-nite-size scaling analysis of the
helicity modulus $\gamma$ as shown in Fig.\ \ref{fig:scaling_gamma}. This value is
very close to $K_c=0.4539\pm 0.0013$ by the series expansion as in
Ref.\ \cite{ferer}, which is also used in Ref.\ \cite{cha1}. This result confirms
the validity of our numerical codes. Using our value of $K_c$, we can also obtain
the universal conductivity $\sigma^{*}/\sigma_{Q}=0.282\pm 0.005$.
This result is also very close to the value
$\sigma^{*}/\sigma_{Q}=0.285\pm 0.02$, obtained in Ref.\ \cite{cha1}.

\subsection{Finite $\Delta A_{ij}$}

\begin{figure}[ht!]
\begin{center}
\includegraphics[width=0.4\textwidth]{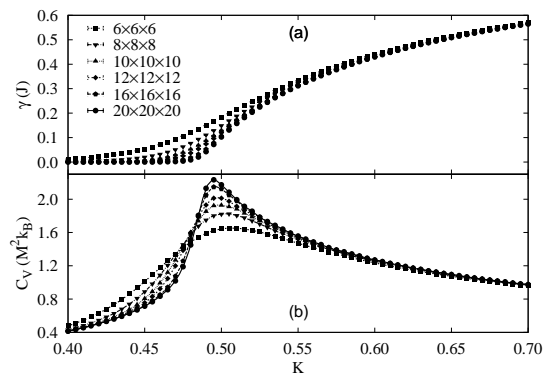}\\%
\end{center}
\caption{\label{fig:gamma_sh01}(a) Helicity modulus $\gamma$ and (b)
specific heat $C_V$, plotted as functions of coupling constant $K$ for several
lattice sizes when $\Delta A_{ij}=1/2$. The error bars in (a), as obtained from
the jackknife method, are smaller than the symbol sizes.
The lines in (b) are cubic spline fits to the Monte Carlo data.}
\end{figure}

\begin{figure}[ht!]
\begin{center}
\includegraphics[width=0.4\textwidth]{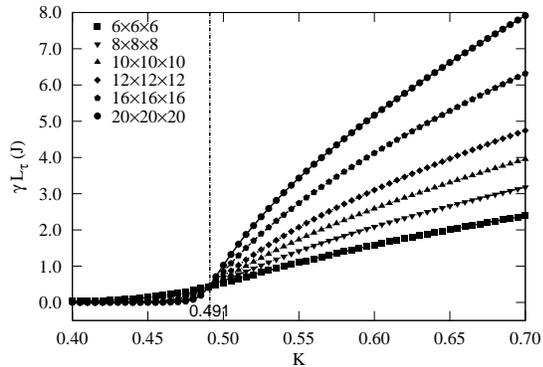}\\%
\end{center}
\caption{\label{fig:scaling_gamma01}Fi\-nite-size scaling behavior of the
data in Fig.\ \ref{fig:gamma_sh01}(a), using $z = 1$.
The apparent crossing point yields $K_c=0.491\pm 0.001$.}
\end{figure}

Figs.\ \ref{fig:gamma_sh01}(a), \ref{fig:gamma_sh01}(b), and
\ref{fig:scaling_gamma01} show the helicity modulus $\gamma$, the specific heat
$C_V$, and the fi\-nite-size scaling behavior of $\gamma$ as a function of coupling
constant $K$ for several lattice sizes when $\Delta A_{ij}=1/2$. When $K>0.55$,
$\gamma$ and $C_V$ appear to be nearly lattice size independent. The error bars from the
jackknife method \cite{newman1} are shown in Fig.\ \ref{fig:gamma_sh01}(a), but
they are smaller than the symbol sizes. The lines are cubic spline fits to the
data in Fig.\ \ref{fig:gamma_sh01}(b). The apparent crossing point in 
Fig.\ \ref{fig:scaling_gamma01} yields the critical coupling constant
$K_c=0.491\pm 0.001$, which is very close to the peak of $C_V$ in
Fig.\ \ref{fig:gamma_sh01}(b).

\begin{figure}[ht!]
\begin{center}
\includegraphics[width=0.4\textwidth]{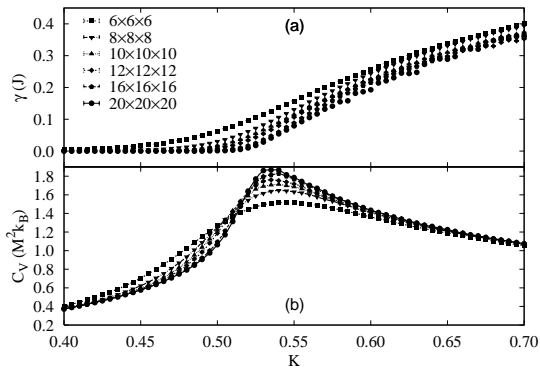}\\%
\end{center}
\caption{\label{fig:gamma_sh02}Same as Fig.\ \ref{fig:gamma_sh01},
except that $\Delta A_{ij}=1/\sqrt{2}$.}
\end{figure}

\begin{figure}[ht!]
\begin{center}
\includegraphics[width=0.4\textwidth]{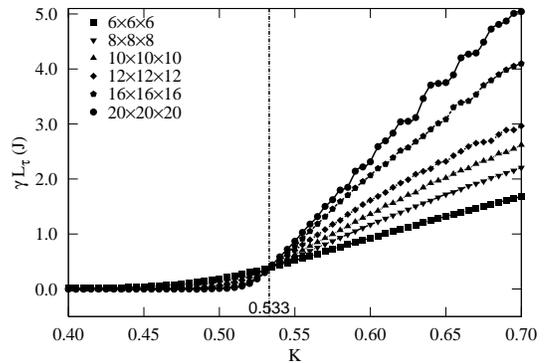}\\%
\end{center}
\caption{\label{fig:scaling_gamma02}Same as Fig.\ \ref{fig:scaling_gamma01},
except that the data are from Fig.\ \ref{fig:gamma_sh02}(a). The apparent crossing
point yields $K_c=0.533\pm 0.001$, as indicated by the vertical dashed line.}
\end{figure}

The corresponding results for $\Delta A_{ij} = 1/\sqrt{2}$ are shown in
Figs.\ \ref{fig:gamma_sh02}(a), \ref{fig:gamma_sh02}(b), and
\ref{fig:scaling_gamma02}. Compared to Fig.\ \ref{fig:gamma_sh01}(a), $\gamma$ shows
more lattice size dependence when $K>K_c$ in Fig.\ \ref{fig:gamma_sh02}(a). The 
apparent crossing point in Fig.\ \ref{fig:scaling_gamma02} yields $K_c=0.533\pm 0.001$.
This $K_c$ is also very close to the peak in $C_V$ as in 
Fig.\ \ref{fig:gamma_sh02}(b).

\begin{figure}[ht!]
\begin{center}
\includegraphics[width=0.4\textwidth]{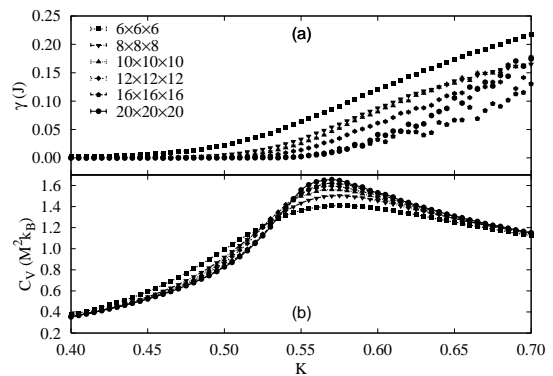}\\%
\end{center}
\caption{\label{fig:gamma_sh03}Same as Fig.\ \ref{fig:gamma_sh01},
except that $\Delta A_{ij}=0.854$.}
\end{figure}

\begin{figure}[ht!]
\begin{center}
\includegraphics[width=0.4\textwidth]{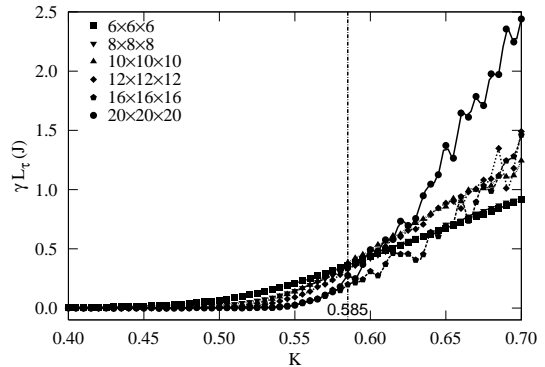}\\%
\end{center}
\caption{\label{fig:scaling_gamma03}Same as Fig.\ \ref{fig:scaling_gamma01},
except that the data are from Fig.\ \ref{fig:gamma_sh03}(a). The apparent crossing 
point yields $K_c=0.585\pm 0.004$.}
\end{figure}

The results for $\gamma$, $C_V$, and $\gamma L_\tau$ when $\Delta A_{ij} = 0.854$
are shown in Figs.\ \ref{fig:gamma_sh03}(a), \ref{fig:gamma_sh03}(b), and
\ref{fig:scaling_gamma03}. In this case, the lattice size-de\-pend\-ence of
$\gamma$ when $K>K_c$ in Fig.\ \ref{fig:gamma_sh03}(a) becomes more conspicuous
than that of Fig.\ \ref{fig:gamma_sh02}(a). The apparent crossing point for the 
different sizes in Fig.\ \ref{fig:scaling_gamma03} is less clearly defined than
in the previous examples, but yields $K_c=0.585\pm 0.004$. This $K_c$ is slightly
larger than the value of $K$ at the maximum of the broad peak in $C_V$, as in
Fig.\ \ref{fig:gamma_sh03}(b). There are some fluctuations of $\gamma L_\tau$
around $K=0.70$ for the lattice size of $12^3$ and larger fluctuations above $K_c$ 
for the lattice sizes of $16^3$ and $20^3$ in Fig.\ \ref{fig:scaling_gamma03}.

The fact that the apparent crossing point in Fig.\ \ref{fig:scaling_gamma03} is 
even less clear than those for smaller values of $\Delta A_{ij}$ suggests that
$z \neq 1$.  We present a more likely scenario for this and other values of 
$\Delta A_{ij}$ in the discussion section below.

\begin{figure}[ht!]
\begin{center}
\includegraphics[width=0.4\textwidth]{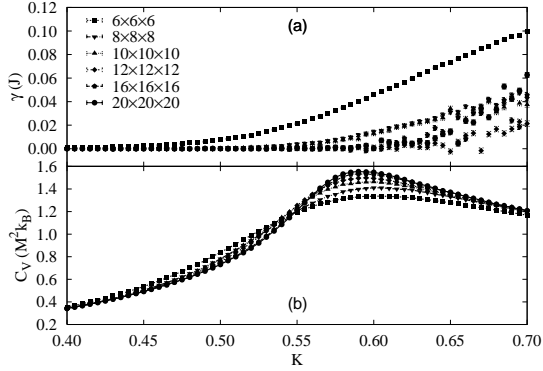}\\%
\end{center}
\caption{\label{fig:gamma_sh04}Same as Fig.\ \ref{fig:gamma_sh01},
except that $\Delta A_{ij}=1.0$.}
\end{figure}

\begin{figure}[ht!]
\begin{center}
\includegraphics[width=0.4\textwidth]{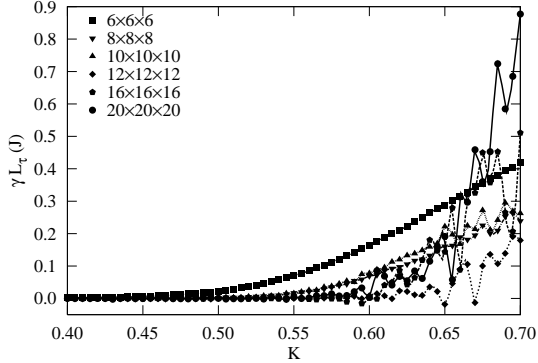}\\%
\end{center}
\caption{\label{fig:scaling_gamma04}Same as Fig.\ \ref{fig:scaling_gamma01},
except that the data are from Fig.\ \ref{fig:gamma_sh04}(a). In this case,
the plots of $L_\tau\gamma(K)$ for different $N_x$ do not cross, suggesting that
the helicity modulus $\gamma$ is no longer a suitable order parameter at
$\Delta A_{ij}=1.0$.}
\end{figure}

As a final calculation for partially random $A_{ij}$, we use $\Delta A_{ij}=1.0$.
The corresponding three thermodynamic variables $\gamma$, $C_V$, and
$\gamma L_\tau$ are shown in Figs.\ \ref{fig:gamma_sh04}(a),
\ref{fig:gamma_sh04}(b), and \ref{fig:scaling_gamma04}, respectively. The lattice
size-de\-pend\-ence of $\gamma$ in Fig.\ \ref{fig:gamma_sh04}(a) becomes far more
conspicuous than the previous two examples. The peak of $C_V$ is very broad, as
shown in Fig.\ \ref{fig:gamma_sh04}(b). In addition, there is nothing like a clear 
crossing point of $\gamma L_\tau$ for different sizes $N_x$ in
Fig.\ \ref{fig:scaling_gamma04}. We interpret this result to mean that the
helicity modulus $\gamma$ does not play the role of an order parameter and
that the transition is not a su\-per\-con\-duc\-tor-to-in\-su\-la\-tor
transition of the same character as at $\Delta A_{ij} = 0$. Furthermore, there are 
strong fluctuations of $\gamma L_\tau$ as a function of $N_x$ when $K\geq 0.64$ for 
most lattice sizes, as can be seen in Fig.\ \ref{fig:scaling_gamma04}. We believe 
that, for this value (and, in fact, at all nonzero values) of $\Delta A_{ij}$, this 
is a transition from a Bose glass to a Mott insulator.


\begin{figure}[ht!]
\begin{center}
\includegraphics[width=0.4\textwidth]{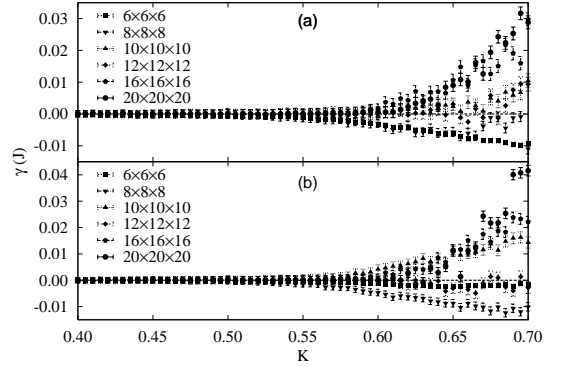}\\%
\end{center}
\caption{\label{fig:gamma_avg}(a) The helicity modulus $\gamma$ as a function
of coupling constant $K$ for several lattice sizes and $\Delta A_{ij}=\infty$.
(b) Same as (a) except that we assume a uniform transverse magnetic field with
frustration $f = 20$ and disorder in the grain positions with a uniformly
distributed random displacement of each site, as described in the text. In both
(a) and (b) we find a negative $\gamma$ for certain values of $K$, depending on
the lattice size. The similarity of (b) and (a) is evidence that these two models
give very similar results.}
\end{figure}

\begin{figure}[ht!]
\begin{center}
\includegraphics[width=0.4\textwidth]{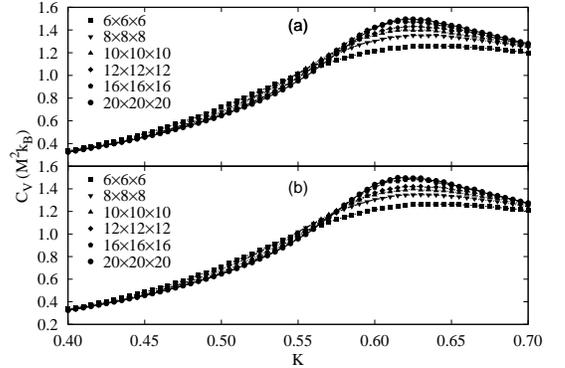}\\%
\end{center}
\caption{\label{fig:sh_avg}Same as Fig.\ \ref{fig:gamma_avg} but for the
specific heat $C_V$. The lines are cubic spline fits to the data.}
\end{figure}

Finally, we have considered the case of a fully random $A_{ij}$,
$\Delta A_{ij}=\infty$. We implemented this by choosing $\Delta A_{ij}$ randomly
between $0$ and $2\pi$. The helicity modulus $\gamma$ and the specific heat $C_V$
for this case are shown in Figs.\ \ref{fig:gamma_avg}(a) and \ref{fig:sh_avg}(a),
respectively. The magnitude of $\gamma$ becomes much smaller than those of
previous cases, so that the error bars are easily visible on the scale of the plot.
The helicity modulus even seems to have negative values for certain values of $K$,
depending on the lattice size. Such negative values and fluctuations of the
helicity modulus in a disordered superconductor were already reported in other
work \cite{spivak}, in the context of a different model.
As in Figs.\ \ref{fig:gamma_sh04}(a) and \ref{fig:scaling_gamma04}, $\gamma$ is
strongly lat\-tice-size dependent and there exists no value of $K$ at which the
curves of $L_\tau \gamma(K)$ for different $N_x$ all cross (we do not show a plot
exhibiting this lack of crossing). All these results indicate that we need a
different order parameter to describe the phase transition. Besides these results, 
we find that the peak in $C_V(K)$ is even broader than that in 
Fig.\ \ref{fig:gamma_sh04}(b). Moreover, the peak of $C_V$ shifts towards a larger 
value of $K$ as $\Delta A_{ij}$ increases.


As a comparison to the fully random $\Delta A_{ij}$ case, we have also considered 
a model of ``positionally disordered sites'' in a strong uniform transverse magnetic
field ${\bf B} = B{\bf \hat{z}}$, similar to a model considered in
Ref.\ \cite{shih2}. The position coordinates $(x_i,y_i)$ of each site are assumed
uniformly and independently distributed between $-\Delta$ and $\Delta$ with respect
to the position $(x_{i0},y_{i0})$ the site would have in the ordered lattice, i.e.,
\begin{eqnarray}
|x_i - x_{i0}| & \leq & \Delta, \nonumber \\
|y_i - y_{i0}| & \leq & \Delta.
\label{eq:posdis}
\end{eqnarray}
In our calculations, we have chosen $\Delta=a/4$, where $a$ is the lattice
constant of the unperturbed lattice. Thus $A_{ij}$ has the form
\begin{equation}
A_{ij}=\frac{2\pi}{\Phi_0}B\frac{x_i+x_j}{2}(y_j-y_i)
\label{eq:aij}
\end{equation}
for near\-est-neigh\-bor sites $i$ and $j$. 
In order to consider a strong field, we choose $f=Ba^2/\Phi_0=20$. The results for
this system of positionally disordered sites are shown in
Figs.\ \ref{fig:gamma_avg}(b) and \ref{fig:sh_avg}(b). They are qualitatively
similar to those with $\Delta A_{ij}=\infty$, and even quantitatively similar for
$C_V$. We conclude that the model of positionally disordered sites is nearly
equivalent to that with random $A_{ij}$, at least for $\Delta A_{ij} = \infty$.

\begin{figure}[ht!]
\begin{center}
\includegraphics[width=0.4\textwidth]{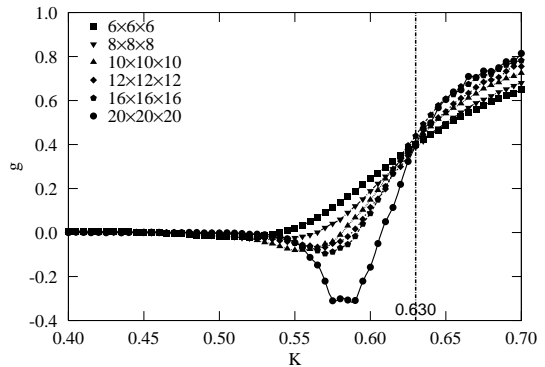}\\%
\end{center}
\caption{\label{fig:g_avg}The renormalized coupling constant $g$
[Eq.\ (\ref{eq:g})] as a function of coupling constant $K$ for several lattice
sizes when $\Delta A_{ij}=\infty$. The crossing point yields $K_c=0.630\pm 0.005$.
The lines are cubic spline fits to the data.}
\end{figure}

At large values of $\Delta A_{ij}$, our results suggest that the transition occurs
between a Mott insulator and a Bose glass rather than a conventional
superconductor. Since a new order parameter is demanded to study this transition,
we use the ``renormalized coupling constant'' $g$ as in Ref.\ \cite{huse}. Using
the same $A_{ij}$ for each realization, two replicas of phase $\theta_j$ are
simulated with different initial conditions and updated using different random
numbers. Their overlap is calculated from the quantity
\begin{equation}
q=\sum_j \exp[i(\theta_j^{(1)}-\theta_j^{(2)})],
\label{eq:overlap}
\end{equation}
where $\theta_j^{(1)}$ and $\theta_j^{(2)} $ are the phases at site $j$ in the
two replicas. Given $q$, the renormalized coupling constant $g$ is defined as
\begin{equation}
g=2-\frac{[\langle|q|^4\rangle]}{[\langle|q|^2\rangle]^2},
\label{eq:g}
\end{equation}
where $\langle\cdots\rangle$ denotes the thermal average while $[\cdots]$ denotes
an average over many realizations of $A_{ij}$.
Figure~\ref{fig:g_avg} shows this $g$ as a function of coupling constant $K$ for
several lattice sizes when $\Delta A_{ij}=\infty$. From the crossing point for
different sizes, we obtain $K_c=0.630\pm 0.005$. Unlike the results in
Ref.\ \cite{huse}, $g$ still has a size dependence when $K>K_c$. 

Figs.\ \ref{fig:gamma_avg}--\ref{fig:g_avg} strongly suggest that 
$\Delta A_{ij} = \infty$ corresponds to a Bose glass transition, rather than a 
conventional superconducting transition. Hence, we expect $z \neq 1$. In order to 
allow for $z \neq 1$, we 
have carried out additional calculations, using a method suggested by 
Guo {\it et al.}\ \cite{guo} and by Rieger and Young \cite{rieger}.  Following the 
procedure of these authors, we first calculate $g$ as a function of the time 
dimension $L_\tau$ for various sizes $L$ and several temperatures $T$.  Since the 
proper scaling behavior of $g$ is not expected to depend on the anisotropy of the 
coupling constants, we assume the same coupling constant $J = 1$ in both the space 
and imaginary time directions.  For each $T$ and $L$, $g$ has a maximum value as a 
function of $L_\tau$.  According to Refs.\ \cite{guo} and \cite{rieger}, the true 
$T_c$ is the temperature such that this maximum value, $g_{\mathrm{max}}$, is 
independent of $L$.  Once $T_c$ is determined by this procedure, the correct $z$ is 
that value which causes a plot of $g(T_c, L_\tau/L^z)$ versus the scaling variable 
$L_\tau/L^z$ to be independent of $L$.  

\begin{figure}[ht!]
\begin{center}
\includegraphics[width=0.4\textwidth]{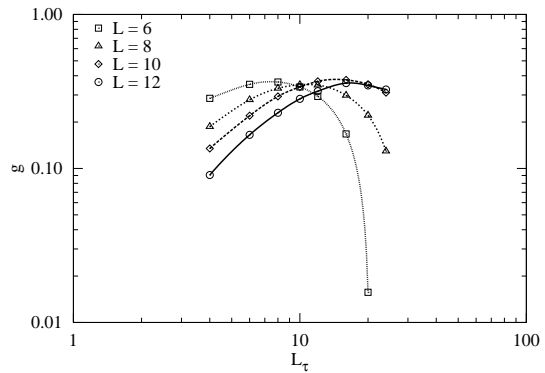}\\%
\end{center}
\caption{\label{fig:guorieger01}Plot of $g(L_\tau, L, T)$ versus $L_\tau$ for 
several values of $L$, as given in the legend, for $T = 1.61J/k_B$ 
and $\Delta A_{ij} = \infty$. In these calculations, the coupling constants $J$ and 
$J_\tau$ are each taken to be unity.  For each $L$, each calculation represents an 
average over 100 realizations of the disorder.  The maximum values 
$g_{\mathrm{max}}(L)$ are nearly independent of $L$.}
\end{figure}

Following this prescription, we have calculated $g(L_\tau, L, T)$ as a function of 
$L_\tau$ for various values of $T$ and a lattice of size $L \times L \times L_\tau$, 
assuming that $J = J_\tau = 1$.  We find that $g_{\mathrm{max}}$ is most nearly 
independent of $L$ when $T = 1.61 J/k_B$.  To illustrate this independence, we plot 
$g(L_\tau, T = 1.61J/k_B)$ for several choices of $L$ in Fig.\ \ref{fig:guorieger01}. 
At this temperature, for all $L$ studied, $g(L_\tau, L, T)$ has a maximum of 
around 0.38 when plotted against $L_\tau$.

\begin{figure}[ht!]
\begin{center}
\includegraphics[width=0.4\textwidth]{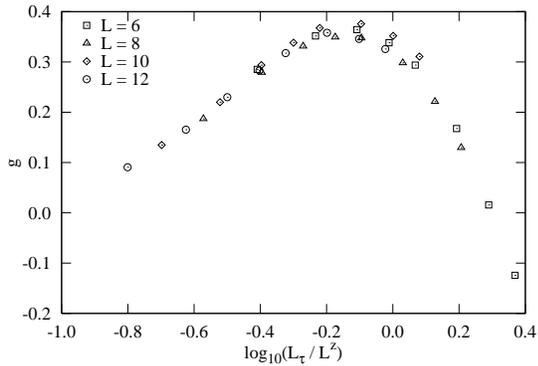}\\%
\end{center}
\caption{\label{fig:guorieger02}Plot of $g(L_\tau, L, T)$ versus 
$\log_{10}(L_\tau/L^z)$ for several values of $L$, as given in the legend, for 
$T = 1.61J/k_B$, $z = 1.3$, and $\Delta A_{ij} = \infty$.  In these calculations, 
the coupling constants $J$ and $J_\tau$ are each taken to be unity.  For each $L$, 
each point represents an average over 100 realizations of the disorder.  For this 
choice of $z$, the results for different values of $L$ collapse very well onto a 
single plot.  The corresponding plots for $z = 1.2$ and $z = 1.4$ produce only 
slightly inferior collapses. We conclude that the correct value of $z$ for this 
transition is $z \sim 1.3 \pm 0.1$.}
\end{figure}

Given $T_c$, we obtain $z$ by plotting $g(T_c, L_\tau/L^z)$ as a function of the 
scaling variable $L_\tau/L^z$ for various values of $L$.  The correct value of $z$ 
is the one which causes these curves to be most nearly independent of $L$. We have 
made such plots for various values of $z$ at $T_c = 1.61J/k_B$, and find that this 
collapse of the numerical data is most nearly obtained for $z = 1.3$, with an 
uncertainty of about $\pm 0.1$. The resulting scaling fit is shown in 
Fig.\ \ref{fig:guorieger02} for $z = 1.3$.  The fit is very good, suggesting that 
(i) the transition for $\Delta A_{ij} = \infty$ is indeed a Bose glass transition, 
and (ii) the critical exponent $z$ at the transition is $z \sim 1.3 \pm 0.1$.

\begin{figure}[ht!]
\begin{center}
\includegraphics[width=0.4\textwidth]{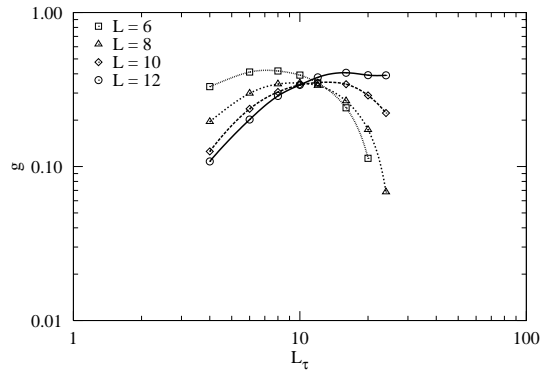}\\%
\end{center}
\caption{\label{fig:guorieger03}Same as Fig.\ \ref{fig:guorieger01} except for 
$\Delta A_{ij} = 1.0$ and $T = 1.70 J/k_B$.}
\end{figure}

\begin{figure}[ht!]
\begin{center}
\includegraphics[width=0.4\textwidth]{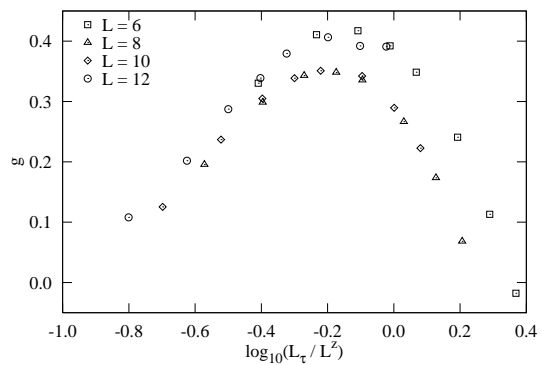}\\%
\end{center}
\caption{\label{fig:guorieger04}Same as Fig.\ \ref{fig:guorieger02} except for 
$\Delta A_{ij} = 1.0$ and $T = 1.70 J/k_B$.}
\end{figure}

We have carried out a similar series of calculations at $\Delta A_{ij} = 1.0$.  For 
this choice, the best glass scaling fits are reasonable, but not so good as for 
$\Delta A_{ij} = \infty$.  They are shown in Figs.\ \ref{fig:guorieger03} and 
\ref{fig:guorieger04} for $T = 1.70J/k_B$, which is our best estimate for the 
glass transition temperature of this model at $\Delta A_{ij} = 1.0$.  Our conclusion 
is that, for $\Delta A_{ij} = 1.0$, the sizes we can achieve ($L \sim L_\tau \sim 12$) 
are simply not large enough to reveal the excellent scaling behavior which is 
expected for a sufficiently large sample.  We discuss below a possible explanation 
why $\Delta A_{ij} = 1.0$ requires a larger sample size than 
$\Delta A_{ij} = \infty$. 


\begin{figure}[ht!]
\begin{center}
\includegraphics[width=0.4\textwidth]{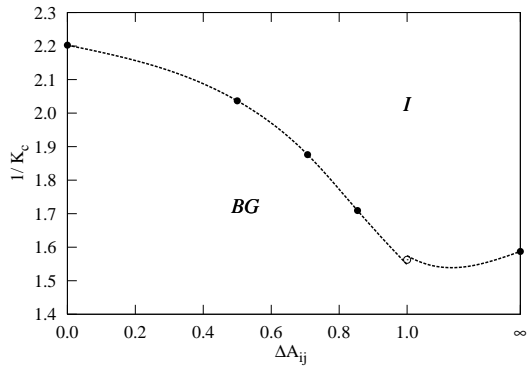}\\%
\end{center}
\caption{\label{fig:invkc}Calculated inverse critical coupling constant $1/K_c$
as a function of $\Delta A_{ij}$. The filled points denote the calculated points,
and the dashed line connecting them is a freehand interpolation of the
data.  We denote the entire ordered region for $\Delta A_{ij} \neq 0$ as ``BG,'' 
consistent with what we believe to be the most probable nature of the ordered 
state.  For $\Delta A_{ij} \leq 0.854$, the data come from calculations of the 
helicity modulus, as described in the text; for $\Delta A_{ij} = \infty$, they 
come from calculation of the glass order parameter $g$, and for 
$\Delta A_{ij} = 1.0$, they come from both, as shown in the Figure.}
\end{figure}

With all the $K_c$'s we have collected so far, we can plot $1/K_c$ as a function
of $\Delta A_{ij}$. This is shown in Fig.\ \ref{fig:invkc}. 
Since $K_c = \sqrt{[J/(2U)]_c}$ and since $1/K_c$ decreases as $\Delta A_{ij}$
increases, these results mean that $[J/(2U)]_c$ {\em increases} with increasing
$\Delta A_{ij}$. Therefore, there exist certain values of the ratio $J/U$ such
that the system is superconducting (or in a Bose glass state) for small 
$\Delta A_{ij}$, but insulating for large $\Delta A_{ij}$. 
As discussed earlier, an increasing value of $\Delta A_{ij}$ can be interpreted
as an increasing value of magnetic field $f\Phi_0/a^2$ for a slightly disordered
Josephson junction array in a transverse magnetic field equal, on average, to an
integer number $f$ of flux quanta per plaquette. Thus, our results suggest that,
for certain values of $J/U$ and integer $f$, the system undergoes a
superconductor (or Bose glass) to insulator transition as $f$ increases. Since
a given array would be expected to have a fixed value of $J/U$, such an array may
undergo an S-I (or BG-I) transition as a function of integer $f$ if $J/U$ is in the
appropriate range. Our results may not be directly applicable to a realistic
thin superconducting film in a magnetic field because such a film is unlikely to
have the topology of a Josephson junction network. However, the two could exhibit
similar phase diagrams.


In Fig.\ \ref{fig:invkc}, we have shown a possible phase diagram for this system,
based on all the numerical data we have accumulated.  We have drawn the diagram
to suggest that the entire ordered region for $\Delta A_{ij} \neq 0$ is of the Bose
glass type, rather than the S type with $z = 1$. This point is discussed further in
the next section.  Despite this assumption, we have, for 
$\Delta A_{ij} \leq 0.854$, obtained $K_c$ from our calculations of the helicity 
modulus, as discussed above.  For $\Delta A_{ij} = \infty$, we have used our 
scaling calculations based on the glass order parameter $g$.  For 
$\Delta A_{ij} = 1.0$, we used both methods.  They give slightly different values 
of $K_c$ at the phase boundary. It is conceivable (but, we believe, unlikely) that 
there is another phase boundary separating the S and BG regions somewhere 
around $\Delta A_{ij} = 1.0$.  Our reasons for believing this scenario to be 
unlikely are given in the discussion below.

\subsection{Conductivity}

If the transition in our model is from a Mott insulator (I) to a superconductor 
(S), then the helicity modulus $\gamma$ is finite in the S state but vanishes 
in the state I. Precisely at the transition, $\gamma$ becomes linear in 
frequency, and the conductivity at the transition can be extracted by a scaling 
analysis \cite{cha1}, as we review below. 
In what follows, we carry out the scaling analysis over the full range of 
$\Delta A_{ij}$, whether the ordered state is S or BG.


In order to obtain the value of the conductivity at the transition, we need the
generalization of the scaling formulas to fre\-quen\-cy-de\-pend\-ent $\gamma$.
When there is such a frequency dependence, Eq.\ (\ref{eq:gamma1}) is generalized
to \cite{cha1}
\begin{equation}
KM\gamma(k)=\tilde{G}(M^{1/\nu}\delta, kM),
\label{eq:gammak}
\end{equation}
where $k=2\pi n/M$ and $n$ is an integer. 
Precisely at $K=K_c$, $\tilde{G}$ will be a function of only $kM$, since
$K - K_c = 0$. Thus we can introduce another scaling function $P$, in terms of
which Eq.\ (\ref{eq:gammak}) can be simplified to
\begin{equation}
KM\gamma(k)=P(kM).
\label{eq:gammap}
\end{equation}
From Eq.\ (\ref{eq:conduct}), the conductivity is obtained by taking the limit
$k\rightarrow 0$ after first taking the limit $M\rightarrow \infty$ with a small
$k$ \cite{cha1}, so that $P(kM)\simeq kM$ in the limit $M\rightarrow \infty$.
Using the scaling function $P$, Eq.\ (\ref{eq:conduct}) can be rewritten
as \cite{cha1}
\begin{equation}
\frac{\sigma^{*}}{\sigma_{Q}}=2\pi \lim\limits_{kM\to \infty}\frac{P(kM)}{kM},
\label{eq:sigma}
\end{equation}
where we have also used the relation $\rho(k) = K\gamma(k)$.
Since this quantity is to be calculated for $k \rightarrow 0$ after
$M \rightarrow \infty$, the ratio $\sigma^{*}/\sigma_Q$ will be finite only if
the scaling function $P(x) \propto x$ in this regime. Since $k=2\pi n/M$, 
the scaling form (\ref{eq:sigma}) can be written again as \cite{cha1}
\begin{equation}
\frac{\sigma(n)}{\sigma_{Q}}=\frac{P(2\pi n)}{n}.
\label{eq:sigman}
\end{equation}

This scaling form is expected to be valid only in the regime $1 \ll n \ll M$.
Since it is difficult to carry out calculations for $M$ large enough that these
inequalities are satisfied, especially for a disordered system, it is necessary to
incorporate corrections to scaling and express $\sigma$ in terms of $n$ and $M$
separately. Since the corrections to scaling vanish in the limit
$n/M \rightarrow \infty$, we expand $\sigma(n)$ as a function of $n$ and $M/n$ using
the same form assumed in Ref.\ \cite{cha1}, namely
\begin{equation}
\frac{\sigma(n,M/n)}{\sigma_{Q}}=\frac{\sigma^{*}}{\sigma_{Q}}+d\left (
\frac{\alpha}{n}-\frac{n}{M}\right ) + \cdots,
\label{eq:sigmanl}
\end{equation}
where $d$ and $\alpha$ are fitting constants. The universal conductivity
$\sigma^{*}$ is found by plotting $\sigma(n,M/n)$ as a function of the scaling
variable $(\alpha/n-n/M)$ for several lattice sizes $M$ and finding the optimal
value of $\alpha$ which produces the best data collapse onto a single curve. The
universal conductivity for this value of $\Delta A_{ij}$ is the value of
$\sigma^*$ at which $\alpha/n - n/M = 0$.

\begin{figure}[ht!]
\begin{center}
\includegraphics[width=0.4\textwidth]{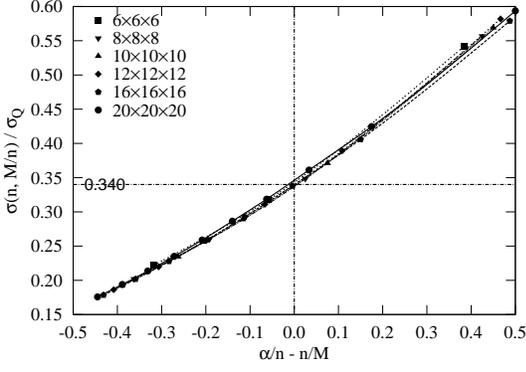}\\%
\end{center}
\caption{\label{fig:conduct_fit01}The conductivity $\sigma(n,M/n)$ divided by
$\sigma_{Q}$ as a function of the variable $\alpha/n-n/M$ for several
lattice sizes $M$ when $\Delta A_{ij}=1/2$ and $K = K_c=0.491$.
The optimal $\alpha$ used here is $0.55$. The universal conductivity
$\sigma^*/\sigma_Q$ is given by that value of $\sigma(n,M/n)$ for which
$\alpha/n-n/M = 0$, as indicated by the vertical dashed line. The
universal conductivity thus obtained is $\sigma^{*}/\sigma_{Q}=0.340\pm 0.006$.}
\end{figure}

\begin{figure}[ht!]
\begin{center}
\includegraphics[width=0.4\textwidth]{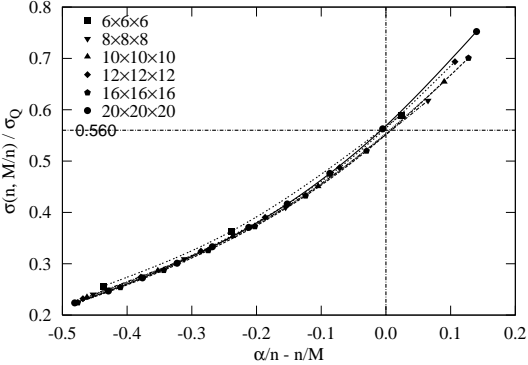}\\%
\end{center}
\caption{\label{fig:conduct_fit02}Same as Fig.\ \ref{fig:conduct_fit01},
except that $\Delta A_{ij}=1/\sqrt{2}$ and $K_c=0.533$. The optimal $\alpha$
used here is $0.19$. The universal conductivity is
$\sigma^{*}/\sigma_{Q}=0.560\pm 0.009$.}
\end{figure}

\begin{figure}[ht!]
\begin{center}
\includegraphics[width=0.4\textwidth]{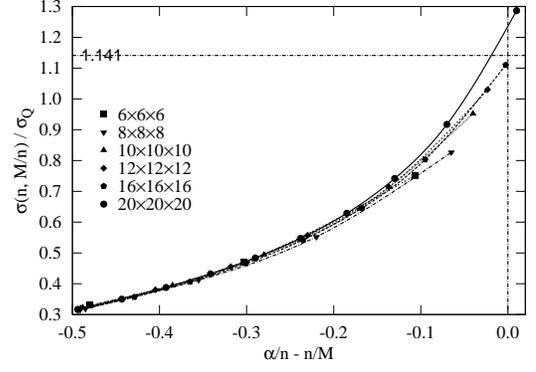}\\%
\end{center}
\caption{\label{fig:conduct_fit03}Same as Fig.\ \ref{fig:conduct_fit01},
except that $\Delta A_{ij}=0.854$ and $K_c=0.585$. The optimal $\alpha$
used here is $0.06$. We therefore have to extrapolate the plot of the
conductivity $\sigma(n,M/n)/\sigma_{Q}$ to reach the point at
$\alpha/n-n/M = 0$. The universal conductivity thus obtained is
$\sigma^{*}/\sigma_{Q}=1.141\pm 0.088$.}
\end{figure}

\begin{figure}[ht!]
\begin{center}
\includegraphics[width=0.4\textwidth]{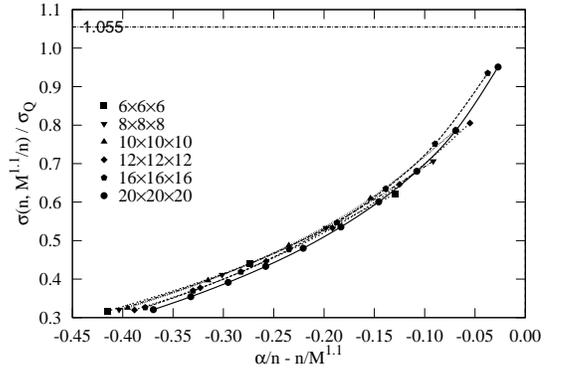}\\%
\end{center}
\caption{\label{fig:conduct_fit04}Same as Fig.\ \ref{fig:conduct_fit03},
except that $\Delta A_{ij}=\infty$, $K_c=0.630$, and $z=1.1$. The optimal $\alpha$
used here is $0.01$. The universal conductivity is also obtained from extrapolation
of the data, resulting in $\sigma^{*}/\sigma_{Q}=1.055\pm 0.090$.}
\end{figure}

Using this method, we find the following universal conductivities for different
values of $\Delta A_{ij}$:
$\sigma^{*}/\sigma_{Q}=0.340\pm 0.006$ when $\Delta A_{ij}=1/2$,
$\sigma^{*}/\sigma_{Q}=0.560\pm 0.009$ when $\Delta A_{ij}=1/\sqrt{2}$,
$\sigma^{*}/\sigma_{Q}=1.141\pm 0.088$ when $\Delta A_{ij}=0.854$, and
$\sigma^{*}/\sigma_{Q}=1.055\pm 0.090$ when $\Delta A_{ij}=\infty$. At each
of these values of $\Delta A_{ij}$, we apply the method just described to
calculate the universal conductivity at the corresponding $K_c$ values obtained
earlier. The results are shown in
Figs.\ \ref{fig:conduct_fit01}, \ref{fig:conduct_fit02}, \ref{fig:conduct_fit03},
and \ref{fig:conduct_fit04}, respectively. The optimal values of $\alpha$'s which
yield these universal conductivities are $\alpha=0.55$, $0.19$, $0.06$, and $0.01$
for $\Delta A_{ij}=1/2$, $\Delta A_{ij} = 1/\sqrt{2}$, $\Delta A_{ij} = 0.854$,
and $\Delta A_{ij} = \infty$, respectively. The accuracy of the calculated
$\sigma^{*}/\sigma_{Q}$ becomes progressively worse as $\Delta A_{ij}$ increases.
In fact, we need to obtain the results for $\Delta A_{ij}=0.854$ and
$\Delta A_{ij}=\infty$ by extrapolation of $\sigma(n,M/n)/\sigma_{Q}$ to the
optimal values of $\alpha$, using Figs.\ \ref{fig:conduct_fit03} and
\ref{fig:conduct_fit04}.

\begin{figure}[ht!]
\begin{center}
\includegraphics[width=0.4\textwidth]{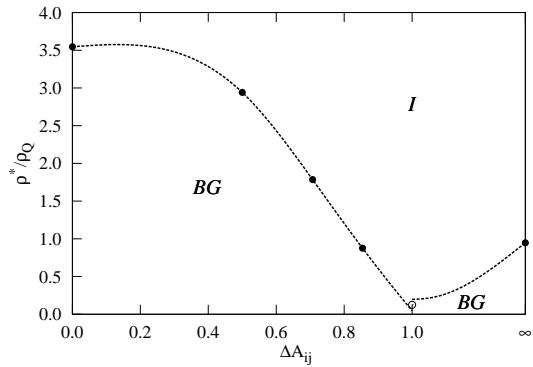}\\%
\end{center}
\caption{\label{fig:rho}The universal resistivity $\rho^{*}$ divided by
$\rho_{Q}$ as a function of $\Delta A_{ij}$. In each case
$\rho^*/\rho_Q = \sigma_Q/\sigma^*$. The dashed lines are cubic
spline fits to the data. 
 ``I'' and ``BG'' denote the 
insulating and Bose glass phases, respectively.  The superconducting 
phase, in our interpretation, occurs only at $\Delta A_{ij} = 0$.  The 
open circle at $\Delta A_{ij} = 1.0$ is an average between the two values 
of $\rho^*/\rho_Q$ from the two dashed lines as in Fig.\ \ref{fig:invkc}.}
\end{figure}

The critical coupling constant $K_c$ increases monotonically with increasing
$\Delta A_{ij}$, as we have already noted.  Similarly, the universal
conductivity $\sigma^{*}$ also appears to increase monotonically with 
$\Delta A_{ij}$.  
From these universal conductivities, we can plot the universal resistivities
as a function of $\Delta A_{ij}$. These are shown in Fig.\ \ref{fig:rho}.
The resistivity decreases with increasing $\Delta A_{ij}$ all along the phase 
boundary between the Mott insulator and the phase-or\-dered state.  The points 
represent the results obtained from the QMC simulations, while the dashed lines 
represent a guide to the eye.  


As noted earlier, increasing $\Delta A_{ij}$ in our model corresponds to
increasing magnetic field for a slightly disordered Josephson junction array in a
uniform transverse magnetic field. Thus, our results which show a decrease in the
universal resistivity with increasing $\Delta A_{ij}$, probably cannot be directly
compared to the experiments reported in Refs.\ \cite{markovic1,markovic2} and the
numerical results of Ref.\ \cite{nishiyama}. The experiments consider a disordered
Bi film in a uniform magnetic field rather than a slightly disordered Josephson
junction array in a periodic field. However, both the experiments and our
calculations find a ze\-ro-tem\-per\-a\-ture mag\-net\-ic-field-tuned 
transition from a phase-or\-dered state to an insulator, and both the experimental papers
and previous calculations interpret this phase transition using scaling fits
of the low-tem\-per\-a\-ture transport properties near the critical field, as we do
here for our model.

\section{\label{sec:level4}DISCUSSION}

In this paper, we have calculated the transition between the superconducting and
the insulating state for a model disordered 2D superconductor in a magnetic field.
We treat the superconductor as a square Josephson junction array with an 
intergranular Josephson coupling energy $J$ and a finite capacitive energy 
described by an on-site charging energy $U$. To include the effect of both disorder 
and a transverse magnetic field, we include a random magnetic phase factor $A_{ij}$ 
in the Josephson coupling between grains; $A_{ij}$ is assumed to have a Gaussian 
distribution with zero mean and a root-mean-square width $\Delta A_{ij}$.

Although our model is certainly artificial, it should reasonably represent the 
effects of a magnetic field applied to a disordered 2D array at integer $f$. 
Specifically, the model resembles a {\it spatially disordered} 2D granular 
superconductor in a {\it uniform} magnetic field, in which the plaquettes have 
slightly different areas. For small $\Delta A_{ij}$, the root-mean-square 
frustration per plaquette is small, but at all nonzero $\Delta A_{ij}$, the grain 
plaquettes are randomly frustrated, a feature which should be relevant in real 2D 
films. Our model provides a way of interpolating smoothly between the ze\-ro-field 
and high-field limits \cite{caveat}.  We have confirmed numerically that, at least 
in the high-field limit, the two models give a BG-I transition at the same value 
of $K$.

Our numerical results suggest that, for any value of $\Delta A_{ij}$, the system 
undergoes a transition from an insulating (I) state to an ordered state. 
We believe that the transition is I to Bose glass (BG) over the {\em entire} range of 
$\Delta A_{ij}$, except for $\Delta A_{ij} = 0$.  Supporting this hypothesis is the 
fact that $g$ exhibits excellent scaling at $\Delta A_{ij} = \infty$ and very good 
scaling at $\Delta A_{ij} = 1.0$.  This hypothesis is also indirectly supported by 
the fact that the scaling behavior of $\gamma$ becomes progressively worse as 
$\Delta A_{ij}$ increases.  To provide stronger support for this hypothesis 
numerically for smaller $\Delta A_{ij}$, we would need to go to much larger Monte 
Carlo sample sizes, using the correct value of $z \sim 1.3$.  

In support of this scenario, we now describe a simple argument, based on the 
well-known Harris criterion \cite{harris}, which suggests that the S state is 
unstable against a small random perturbation of the type we consider here. Consider 
the ze\-ro-field version of model (\ref{eq:calh}), in the presence of some kind of weak 
uncorrelated disorder.  In a region of size $\xi^d$ (where $d = 2$ and $\xi$ is the 
correlation length of the unperturbed system), the critical value of $K$ should 
fluctuate by an amount of order $\xi^{-d/2}$.  Near the critical value of $K$, the 
correlation length of the unperturbed system varies with $K$ according to the 
relation $\xi \propto [(K - K_c)/K_c]^{-\nu}$. 
In order for the transition of the unperturbed system to be unaffected by the disorder, 
the Harris criterion suggests that $\nu \geq 2/d$.  For the present case, $d = 2$, but 
because we are dealing with a quantum transition, $\nu$ is that of a 3D $XY$ 
transition, namely $\nu \sim 2/3$ \cite{pelissetto}.  Thus, the inequality
is not satisfied and we expect this quantum phase transition to be unstable against 
point disorder in 2D.

For the present model, the randomness is indeed uncorrelated within the plane, as 
required by the above argument, but it is somewhat different from the usual point 
disorder.  For small $\Delta A_{ij}$, the random part of the Hamiltonian may be 
written
\begin{equation}
{\cal H}^\prime_{\mathrm{ran}}(\Delta A_{ij}) \sim -J\sum_{\langle ij \rangle}
\sin(\theta_i-\theta_j)\Delta A_{ij},
\label{eq:hran}
\end{equation}
where $\Delta A_{ij}$ is a Gaussian random variable.  Despite the form of this disorder, 
it seems reasonable that the disorder would have at least as strong an effect on 
the phase transition of the pure model as more conventional point disorder.  
Therefore, we suggest, based on this rough argument, that the 3D $XY$ phase 
transition of the pure model is unstable against this random field perturbation 
for arbitrarily weak $\Delta A_{ij}$.

The next question is, to what is the 3D $XY$ transition unstable? 
The most likely scenario is that the transition is of the I to BG class over the 
entire range $\Delta A_{ij} \neq 0$.  The seeming presence of the S phase at small 
but finite $\Delta A_{ij}$ is probably due to the fact that our samples are not 
large enough to exhibit the expected BG phase.
For example, at $\Delta A_{ij} = 1/2$, the rms variation in total flux through a 
single plaquette would be $2\cdot(1/2)\Phi_0/(2\pi) = \Phi_0/(2\pi)$.  Thus, the 
rms variation in total flux 
through a lattice of $L^2$ plaquettes would be $L\Phi_0/(2\pi)$ (where the factor
of $L$ comes from the fact that $4L$ is the perimeter of the $L^2$ plaquettes), 
and hence, even for $L = 12$, would be only about two flux quanta.  This value is 
not large enough to yield results characteristic of $\Delta A_{ij} = \infty$, at 
which we have shown the Bose glass is the stable ordered phase. Thus, indeed, the 
sample sizes we have considered are simply not large enough to exhibit fully 
developed Bo\-se-glass scaling at $\Delta A_{ij} \sim 1/2$, or even at larger 
values than this. Nonetheless, we have the basic result that, for any 
$\Delta A_{ij}$, there is a transition from an insulating state to an ordered state 
with increasing values of the coupling parameter $K = \sqrt{J/(2U)}$.  As the above 
argument suggests, we believe that this ordered state is a BG phase for any nonzero 
$\Delta A_{ij}$.

The critical coupling constant $K_c$ for the transition from I to BG increases 
monotonically with increasing $\Delta A_{ij}$. Thus, for certain values of $K$, the 
material is in the BG state at low $\Delta A_{ij}$, but goes through a BG-I 
transition as $\Delta A_{ij}$ increases. In a disordered material, increasing 
$\Delta A_{ij}$ can be identified with increasing transverse magnetic field for a 
slightly disordered Josephson junction array at integer $f$, in the sense which we 
have discussed earlier. Since any given material should have a fixed $K$, 
independent of $\Delta A_{ij}$, this trend implies that some materials, which have 
suitable values of $K$, will go through a BG-I transition with increasing field. A 
material with a smaller $K$ will remain insulating for all $\Delta A_{ij}$, while 
one with a larger $K$, would remain a BG for all fields. All this behavior follows 
from the phase diagram drawn in Fig.\ \ref{fig:invkc}. In each case, the I phase in 
our model is a Mott insulator, since Cooper pairs are localized by Coulomb 
repulsion rather than by disorder.



Besides calculating the critical values of coupling constant $K_c$, we have also
computed the universal conductivities $\sigma^{*}$ as a function of $\Delta A_{ij}$
for these transitions. The values of both $K_c$ and 
$\rho^*/\rho_Q \equiv (\sigma^*/\sigma_Q)^{-1}$ are shown for various values of 
$\Delta A_{ij}$ in Fig.\ \ref{fig:rho}.  
In all cases, these values are obtained by a scaling analysis of the numerically 
calculated helicity modulus $\gamma$ and the renormalized coupling constant $g$. 

Our results may be consistent with experimental findings as in
Refs.\ \cite{hebard,okuma95,yazdani,markovic1,markovic2,gantmakher00,gantmakher03,
baturina04,baturina05,aubin}. The data in these references indicate that 
$a$-InO$_x$ films \cite{hebard,gantmakher00}, granular In films \cite{okuma95},
$a$-MoGe films \cite{yazdani}, Bi films \cite{markovic1,markovic2},
Nd$_{2-x}$Ce$_x$CuO$_{4+y}$ films \cite{gantmakher03},
TiN films \cite{baturina04,baturina05}, and
Nb$_{0.15}$Si$_{0.85}$ films \cite{aubin} show that the resistance per
square, normalized by the resistance at the transition, at very low temperatures 
decreases as a function of the scaled magnetic field $B$ when the magnetic
field is less than the critical value $B_c$, while it increases when $B>B_c$. 
It should be kept in mind, of course, that our model calculations refer to a 
disordered Josephson array at an integer number of flux quanta per plaquette, on 
average, whereas the experiments deal with systems having a possibly different 
topology.



Numerically, there are several ways in which our calculations could be further
improved. In some cases, the number of realizations (100) we have used for $A_{ij}$
may be insufficient to provide accurate statistics and might lead to significant
numerical uncertainties. Our choice for the number of realizations is dictated by
a compromise between computing costs and statistical errors. Because of the large
amount of computing time involved, we have carried out our calculations only up
to a lattice size at most of $20$ on an edge, and have considered only five 
different nonzero $\Delta A_{ij}$'s. Our results would have had greater accuracy
and given a more detailed picture of the phase diagram if we had been able to 
include more values of $\Delta A_{ij}$, a larger number of realizations, and, 
especially, larger lattice sizes.  In addition, our 
``world-line'' algorithm \cite{hirsch,batrouni90,batrouni92} could be replaced by
other approaches, such as a ``worm'' algorithm \cite{fabien} or a stochastic series
expansion \cite{sandvik,hebert,yunoki,smakov}, possibly leading to better 
convergence.  It might also be valuable to develop another model in which the 
disorder is introduced in a manner closely resembling that in actual 
superconducting films.  Finally, we note that the same approach could be used to
calculate the fi\-nite-fre\-quen\-cy conductivity of the low-tem\-per\-a\-ture 
phase for various values of $\Delta A_{ij}$ \cite{runge,sorensen,wallin}.




To summarize, we have carried out extensive quantum Monte Carlo simulations of a 
model for a transition from a Mott insulator to a superconducting phase at low 
temperatures.  The model is characterized by a continuously tunable disorder 
parameter $\Delta A_{ij}$.  Our numerical results suggest that, for any nonzero 
$\Delta A_{ij}$, there is such a transition, and that the ordered phase is a Bose 
glass.  The evidence that the ordered phase is a Bose glass is strong for 
$\Delta A_{ij} = \infty$,  but less conclusive for smaller $\Delta A_{ij}$.  We 
also find that, for certain values of the coupling variable $K$, the system can go 
from BG at small $\Delta A_{ij}$ to a Mott insulator at large $\Delta A_{ij}$.  We 
also discuss the possibility that this transition may be related to the 
field-driv\-en superconductor to insulator transition seen in a number of 
superconducting films.  It would be of great interest if our results could be 
compared to a suitable experimental realization.

\begin{acknowledgments}
This work was supported by NSF Grant No.\ DMR04-13395. All of the calculations
were carried out on the P4 Cluster at the Ohio Supercomputer Center, with the
help of a grant of time. We thank Prof. S.\ Teitel for several valuable discussions.
\end{acknowledgments}


\begin{thebibliography}{99}
\bibitem{caveat} Our model does differ in one respect from a spatially disordered
granular system with a uniform applied magnetic field: for any $\Delta A_{ij}$, the
mean frustration per plaquette is zero.
\end{thebibliography}
\end{document}